\documentclass[sigconf,nonacm]{acmart}
\AtBeginDocument{%
  }

\setcopyright{acmlicensed}
\copyrightyear{2018}
\acmYear{2018}
\acmDOI{XXXXXXX.XXXXXXX}
\acmConference[Conference acronym 'XX]{Make sure to enter the correct
  conference title from your rights confirmation emai}{June 03--05,
  2018}{Woodstock, NY}
\acmISBN{978-1-4503-XXXX-X/18/06}

\usepackage{booktabs}
\usepackage{multirow}
\usepackage{subcaption}

\begin{document}

\title{TAGR: Temporally Adaptive Generative Recommendation for Industrial Live-Streaming Advertising}


\author{Wencai Ye$^{1,*}$, Guangyi Liu$^{1,2,*}$, Chaoyi Wang$^{1}$, Wenbin Luo$^{1}$, Shengyu Wang$^{1}$, Mingjie Sun$^{1}$, Peng Wang$^{1}$, Quanming Yao$^{2}$, Wenjin Wu$^{1,\dag}$, Peng Jiang$^{1}$}
\affiliation{%
  \institution{$^{1}$Kuaishou Technology, Beijing, China \quad $^{2}$Tsinghua University, Beijing, China}
  \city{} 
  \country{}
}
\email{{yewencai, liuguangyi06, wangchaoyi, luowenbin03, wangshengyu, wangpeng16, wuwenjin, jiangpeng}@kuaishou.com}

\thanks{$^{*}$ Equal contribution. $^{\dag}$ Corresponding author. \\
This work was done when G. Liu was an intern at Kuaishou Technology.}

\renewcommand{\shortauthors}{Ye et al.}
\renewcommand{\arraystretch}{1.15}

%

\begin{abstract}
Live-streaming advertising has become an important monetization channel on short-video and e-commerce platforms. Its rapidly evolving live content, promoted products, and user feedback impose stringent freshness requirements on recommendation models. Existing generative recommenders are largely designed for relatively stationary item domains and therefore adapt inadequately at three levels: static semantic IDs (SID) cannot represent evolving live-ad targets, single-scale user behavior modeling cannot reliably capture users' rapidly changing intent, and preference optimization struggles to balance fresh on-policy feedback with training stability.

We propose \textbf{TAGR}, a generative recommendation framework that performs temporal adaptation at three levels: live-ad tokenization, user intent modeling, and preference alignment. At the token level, \textbf{Live Semantic-Collaborative ID (LSID)} periodically updates each active live ad's SID assignment according to its current live scene and promoted products, while retaining a stable hierarchical token vocabulary for autoregressive generation. At the intent level, \textbf{Intent-Aware Generation (IAG)} models live-room entry histories at multiple temporal granularities as the primary intent sequence, preserves auxiliary behavior sequences as separate input channels, and weights next-token prediction (NTP) using post-request \textit{intent evidence} and \textit{business value}. At the alignment level, \textbf{Intermittent On-Policy Preference Optimization (IOPO)} periodically samples fresh candidate groups from the current policy and performs behavior- and value-aligned preference updates interleaved with supervised NTP maintenance to preserve the learned behavior distribution.

Deployed on a large-scale e-commerce live-stream advertising platform, TAGR improves live-room entry and shopping-cart click rates by \textbf{8.5\%} and \textbf{7.4\%}, respectively, and achieves a \textbf{16.1\%} revenue lift in this e-commerce live-stream advertising setting over the production baseline. These results demonstrate the effectiveness and industrial viability of temporally adaptive generative recommendation for live-stream advertising.
\end{abstract}



\maketitle

\section{Introduction}

\begin{figure}[tbp]
        \centering
        \includegraphics[width=1.0\linewidth]{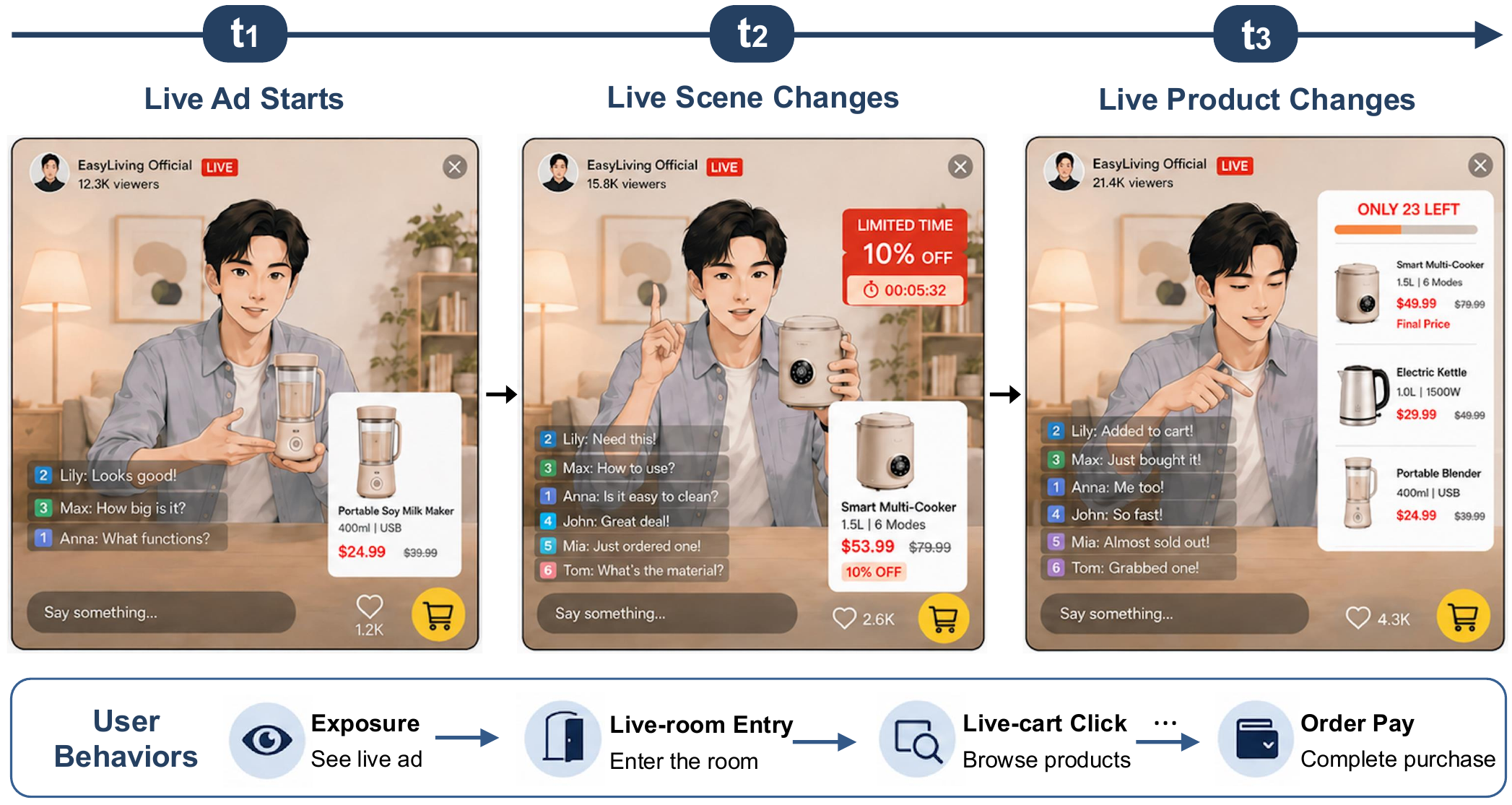}
\vspace{-1em}
    \caption{Live-stream advertising evolves at multiple time scales: live scenes and promoted products reshape live-ad tokens, user actions reveal shifting intent, and post-exposure feedback provides delayed preference signals.}
    \label{fig:background}
\vspace{-4pt}
\end{figure}

Live-streaming advertising has become a major monetization format on short-video and e-commerce platforms, where rapidly changing content and feedback place strong freshness demands on recommendation~\cite{liverec,sliver,momentcross,liveforesighter,tsstfn}. Unlike a conventional ad associated with a relatively stable creative or item, a \textit{live ad} is jointly defined by two time-varying objects: the current live scene and the set of products promoted in that scene. As illustrated in Figure~\ref{fig:background}, the scene captures how the streamer is currently selling, while the product set captures what is being promoted, along with current prices, promotions, and inventories; either the scene or product set may change within minutes, even when the live room remains unchanged. In parallel, a user's intent may evolve from casual viewing to product exploration and purchase as new live content and feedback arrive. Consequently, neither a persistent room identifier nor an individual product identifier alone can represent the serving target. Live-stream advertising must instead retrieve time-sensitive live ads according to the current scene, promoted products, and user intent across multiple time scales.

Industrial advertising systems commonly use multi-stage discriminative pipelines~\cite{din,sim,dlrm}. Their retrieval stages construct candidate sets through fixed indexes and cached representations, after which downstream models progressively estimate engagement and conversion. Generative recommendation provides a different retrieval paradigm by directly producing discrete candidate identifiers from user history and context~\cite{tiger,rqvae,onerec,gr4ad,das}. Its hierarchical output space supports end-to-end retrieval and structured search, and offers a natural interface for representing sparse or newly emerged targets. Recent work further enriches generative recommendation with semantic knowledge and more expressive user conditioning~\cite{letter,eager,lcrec,mmq_v2,align3gr,openonerec}; in live-streaming recommendation, OneLive and SSRLive additionally explore dynamic tokenization and time-aware modeling~\cite{onelive,ssrlive}.

Despite this progress, applying generative recommendations to live-stream advertising presents three temporal challenges. \textbf{First, live token-side temporal adaptation is required as the live ad evolves.} Most non-live generative recommenders use static semantic IDs~\cite{gr4ad,onerec,tiger,das}. Recent live-streaming methods introduce dynamic tokens for evolving live content~\cite{onelive,ssrlive}. However, their targets primarily characterize the live stream or streamer. Live advertising further requires tokens to jointly track the current scene and promoted products while preserving a stable generation space. \textbf{Second, user intent-side temporal adaptation must account for user behavioral scale, behavior type, and supervision reliability.} Conventional sequential generation commonly conditions on a single chronological behavior history~\cite{tiger,onerec,ssrlive}, which can obscure either transient or longer-range intent and lose behavior-specific signals. Moreover, standard next-token learning treats logged targets uniformly~\cite{tiger}, although deeper feedback, such as cart clicks or purchases, provides stronger intent evidence than a brief live-room entry. Intent modeling therefore requires multi-scale, behavior-specific context and reliability-aware supervision. \textbf{Third, preference alignment-side temporal adaptation must balance policy-feedback freshness with behavioral preservation.} Existing off-policy alignment methods learn from lagged-policy candidates~\cite{onerec,gr4ad}, which may cause reward mismatch as the policy and traffic distribution evolve. Continuous online RL improves policy-feedback freshness, but can destabilize training by repeatedly perturbing the behavior distribution learned from supervised NTP. Preference alignment therefore requires current-policy feedback together with a mechanism that preserves supervised behavior. 


These challenges motivate a system-level approach to live-stream advertising that adapts jointly at the token, intent, and alignment levels. \textbf{At the token level, Live Semantic-Collaborative ID (LSID)} learns a stable hierarchical vocabulary through user-scene-product alignment, while periodically refreshing each active live ad's request-independent token assignment according to its current scene and promoted products. \textbf{At the intent level, Intent-Aware Generation (IAG)} models the primary live-room entry sequence at multiple temporal granularities while preserving auxiliary behavior streams and long-term profiles as separate input channels. During supervised training, post-request feedback provides \textit{intent evidence} for how reliably each logged target reflects intent, while user and commercial signals determine its \textit{business value} training priority. \textbf{At the alignment level, Intermittent On-Policy Preference Optimization (IOPO)} periodically learns from candidate groups produced by the current generator and interleaves these updates with supervised maintenance. Its behavior-aligned signal anchors generated targets to observed user feedback, while its value-aligned signal incorporates engagement and commercial feedback. Together, the three components adapt the generation token space, user-intent representation, and preference alignment to the temporal dynamics of live-stream advertising.

To the best of our knowledge, TAGR is the first system-level study of generative recommendation for industrial live-stream advertising. We deploy its generative retrieval stage on a large-scale e-commerce platform and evaluate it through offline retrieval, component analysis, and multi-week production experiments. TAGR improves live-room entry and shopping-cart click rates, as well as cold-start and mid-to-tail coverage, and the full system increases revenue by \textbf{16.1\%} over the production discriminative baseline. These results show that three-level temporal adaptation is both effective and practical under industrial freshness, throughput, and latency constraints.

Our main contributions are summarized as follows:
\begin{itemize}
    \item We formulate live-stream advertising as a generative recommendation problem requiring temporal adaptation at the token, intent, and alignment levels.
    \item We propose TAGR, comprising LSID for dynamic live token construction, Intent-Aware Generation for multi-scale, multi-behavior intent representation, and IOPO for fresh yet stable preference alignment.
    \item We validate TAGR through large-scale offline and production experiments in e-commerce live-stream advertising, obtaining a \textbf{16.1\%} revenue lift together with improved post-exposure engagement, cold-start performance, and mid-to-tail coverage.
\end{itemize}

\begin{figure*}
    \centering
    \includegraphics[width=1.0\linewidth]{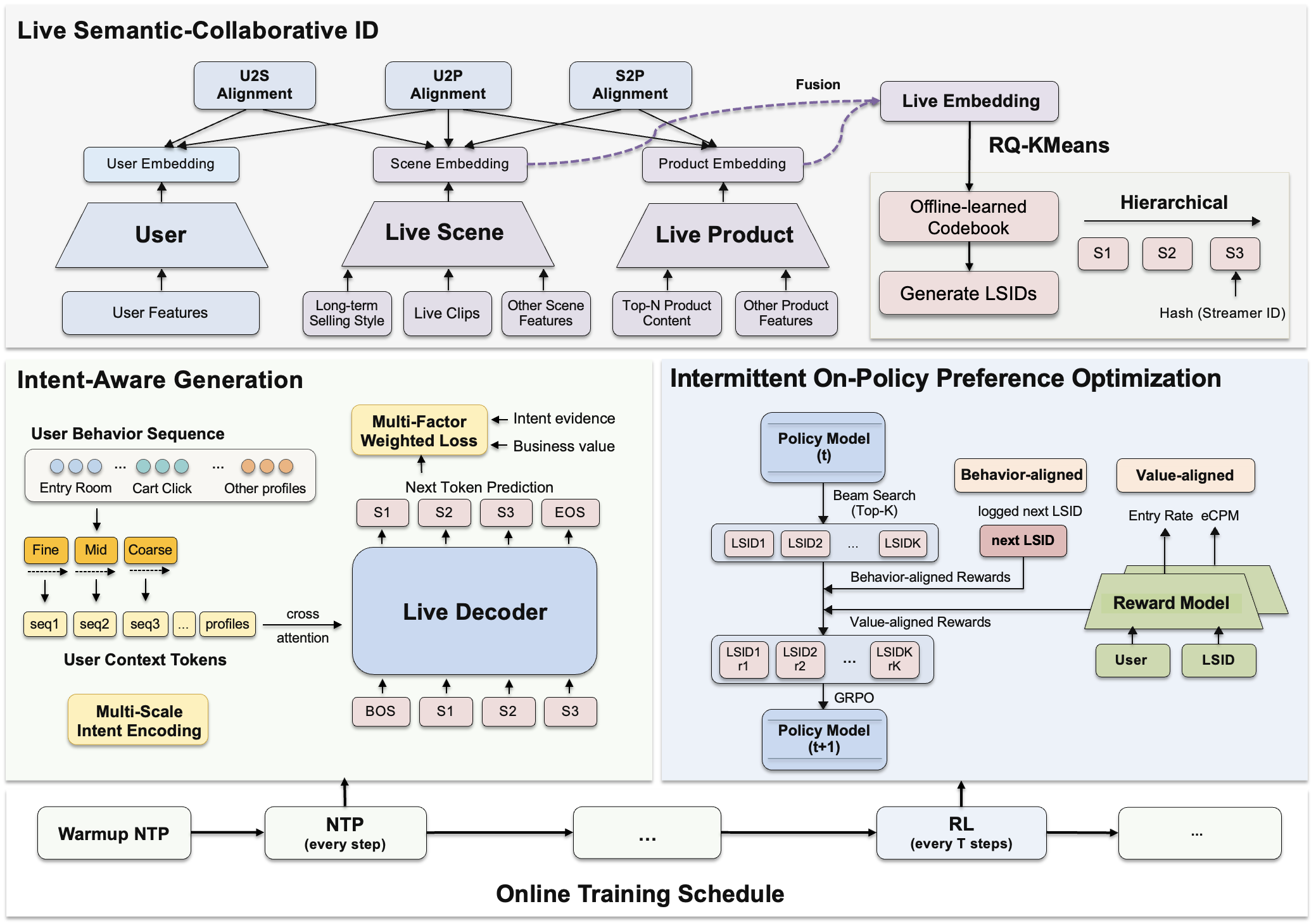}
    \vspace{-6pt}
    \caption{Overview of TAGR, which performs temporal adaptation at three levels: token-side target construction (dynamically updating identifier tokens for live ads), intent-side generation (modeling multi-scale, request-time user intent), and alignment-side optimization (intermittent on-policy preference optimization to align ranking with feedback).}
    \label{fig:TAGR}
\end{figure*}

\section{Methodology}

\subsection{Overview}

We study live-streaming advertising recommendations in an industrial serving scenario. For each user request at time $t$, the system observes a user context $\mathcal{C}_u(t)$, including the user profile, historical behaviors, and recent interactions, together with the available live-stream ads $\mathcal{A}(t)$. Each ad $a \in \mathcal{A}(t)$ is jointly defined by a live scene $S_a(t)$, which describes how the streamer is currently selling, and a promoted-product set $P_a(t)$, which describes the products being promoted and their product-side information at that moment. The objective is to generate a Top-$K_{\text{serve}}$ list that reflects the user's current intent and contains live ads likely to produce consequential engagement and advertising revenue.

TAGR connects the three levels of temporal adaptation in a sequential generation pipeline. LSID first converts each active live ad's current scene and promoted-product set into a time-indexed discrete target $\mathbf{y}_a(t)$; these refreshed assignments define both the supervised labels for generation and the LSID-LiveID index used at serving time. Given these targets, Intent-Aware Generation (IAG) constructs a time-dependent request-time representation $\mathbf{M}_u(t)$ from multi-scale live-room entries, separate auxiliary behavior channels, and long-term profiles, and refreshes $\mathbf{M}_u(t)$ whenever new user behaviors arrive. It then trains the decoder to predict $\mathbf{y}_a(t)$. Within this stage, MF-NTP uses post-request feedback as \textit{intent evidence} and user and commercial signals as \textit{business value} weights, thereby determining how strongly each logged target supervises the base generator. The resulting supervised generator initializes IOPO, which periodically samples fresh candidate groups from the current policy for preference updates and interleaves them with NTP steps that maintain the learned behavior distribution. At serving time, the refined generator produces Top-$K_{\text{serve}}$ LSIDs through beam search, and the same serving index resolves them to active live-stream ads.

\subsection{Live Semantic-Collaborative ID}
\label{subsec:lsid}

At the token level, accurate live-ad retrieval needs an identifier that tracks changes in the live scene and promoted products. A static identifier cannot jointly represent multiple products or distinguish the same product set across live scenes, while a purely semantic identifier misses collaborative evidence from user feedback. We thus propose \textbf{Live Semantic-Collaborative ID (LSID)}, which periodically updates the token assignment of each active live ad based on its latest scene and product-side features. To keep changing assignments learnable, LSID retains a stable hierarchical vocabulary for autoregressive generation and beam search. It learns this target space through user-scene-product alignment while keeping the assignment of each live ad canonical and request-independent.

LSID constructs the live ad token representation from its two defining objects, the scene and product, and uses the user representation to inject collaborative preference information during representation learning. The scene encoder summarizes the streamer's historical selling style, current live-stream content, and real-time scene features; the product encoder summarizes product text, attributes, order status, and product-side features; and the user encoder summarizes long-term interests and profile features. Concretely, $\mathbf{v}_{\text{style}}$ denotes the streamer's long-term selling style extracted from historical product-carrying behaviors, $\mathbf{v}_{\text{clip}}(t)$ denotes the real-time live clip representation, and $\mathbf{f}_S(t)$ denotes other scene-side features. On the product side, $\mathbf{e}_{\text{title}}$ and $\mathbf{e}_{\text{caption}}$ are text embeddings, $\mathbf{a}_{\text{attr}}$ denotes structured product attributes, $\mathbf{o}_{\text{order}}(t)$ captures real-time order status, and $\mathbf{f}_P$ denotes additional product features. For the user, $\mathbf{g}_{\text{interest}}$ and $\mathbf{h}_{\text{profile}}$ denote global interest and profile features. We use three dedicated encoders, implemented as MLPs with ReLU activations~\cite{transformer}, to project these heterogeneous inputs into a shared space:

\begin{equation}
\begin{aligned}
\mathbf{s} &= \mathrm{Enc}_S\!\big([\,\mathbf{v}_{\text{style}};\;
              \mathbf{v}_{\text{clip}}(t);\;
              \mathbf{f}_S(t)\,]\big), \\[2mm]
\mathbf{p} &= \mathrm{Enc}_P\!\big([\,\mathbf{e}_{\text{title}};\;
              \mathbf{e}_{\text{caption}};\;
              \mathbf{a}_{\text{attr}};\;
              \mathbf{o}_{\text{order}}(t);\;
              \mathbf{f}_P\,]\big), \\[2mm]
\mathbf{u} &= \mathrm{Enc}_U\!\big([\,\mathbf{g}_{\text{interest}};\;
              \mathbf{h}_{\text{profile}}\,]\big).
\end{aligned}
\label{eq:tower_def}
\end{equation}

We align the three representations along complementary directions to make the target space both semantically meaningful and behavior-aware. User-to-Scene (U2S) and User-to-Product (U2P) alignments use live-room entry and cart-click signals to encode collaborative preference, while Scene-to-Product (S2P) alignment captures the co-occurrence structure between live scenes and promoted products. We optimize the following contrastive alignment loss~\cite{infonce}:

\begin{equation}
\mathcal{L}_{\text{align}} =
\mathcal{L}_{\text{U2S}}(\mathbf{u},\mathbf{s}) +
\mathcal{L}_{\text{U2P}}(\mathbf{u},\mathbf{p}) +
\mathcal{L}_{\text{S2P}}(\mathbf{s},\mathbf{p}).
\label{eq:align}
\end{equation}

After alignment, we construct the time-dependent live-ad embedding by fusing the current scene and product representations, followed by L2 normalization:

\begin{equation}
\mathbf{z}_{\text{live}}(t) = \mathrm{norm}\big(\alpha\,\mathbf{s}(t) + \beta\,\mathbf{p}(t)\big).
\label{eq:fusion}
\end{equation}

The user representation participates only in contrastive alignment; token assignment itself depends on the scene and product representations in Eq.~\eqref{eq:fusion}. Concretely, collaborative information is injected through the user-to-scene and user-to-product (U2S/U2P) contrastive objectives, which shape the shared embedding space (and hence the quantization/codebook geometry) before request-independent scene--product fusion and token assignment. LSID therefore incorporates collaborative structure without making the serving target specific to the requesting user. We quantize $\mathbf{z}_{\text{live}}(t)$ into a hierarchical token sequence using an RQ-KMeans codebook~\cite{rqvae}, which defines the vocabulary used by the generator. The live-ad embedding is periodically recomputed from the latest scene and product-side features and reassigned against this vocabulary. To reduce collisions without overfitting to individual live streams, the final level uses a streamer-aware hash bucket. Each live-stream ad is thus represented as:

\begin{equation}
\mathbf{y}_a(t) = (s_1, s_2, \dots, s_D), \quad s_\ell \in \mathcal{V}_\ell,
\label{eq:lsid_seq}
\end{equation}
where $s_\ell$ denotes the token at level $\ell$, $D$ is the LSID depth, and $\mathcal{V}_\ell$ is the vocabulary at that level. We omit the request-time index from individual tokens when the context is unambiguous.

LSID thus realizes token-level temporal adaptation by allowing each active live-ad to follow its current scene and promoted products while retaining a consistent hierarchical space for learning and decoding.

\subsection{Intent-Aware Generation}

At the intent level, effective user modeling should capture both the temporal structure of user intent and the different ways in which that intent is expressed, while also distinguishing which logged targets reliably reflect it. We implement \textbf{Intent-Aware Generation (IAG)} with a Lazy Decoder~\cite{onerec_v2}, a lightweight Transformer decoder~\cite{transformer}, and two complementary designs. Multi-Scale Intent Encoding (MSI-Encoding) constructs the time-dependent request-time representation $\mathbf{M}_u(t)$ from a multi-scale live-room-entry sequence, separate auxiliary behavior streams, and long-term profile features. Whenever a new behavior arrives, IAG refreshes $\mathbf{M}_u(t)$ so that the representation evolves with the user's latest evidence. Multi-Factor Weighted Next-Token Prediction (MF-NTP) then uses post-request feedback as \textit{intent evidence} and user- and monetization-side signals as \textit{business value} weights during supervised learning.

Given the user representation $\mathbf{M}_u(t)$ and the previously generated prefix, the decoder autoregressively predicts the LSID sequence $\mathbf{y}=(s_1,\dots,s_D)$:
\[
P_\theta(\mathbf{y}\mid \mathbf{M}_u(t))=\prod_{\ell=1}^{D}P_\theta(s_\ell\mid s_{<\ell},\mathbf{M}_u(t)).
\]
MSI determines how intent is represented across behavioral scales and behavior types at request time, while MF-NTP determines how strongly the corresponding logged target should supervise the generator.

\subsubsection{Multi-Scale Intent Encoding}
\label{subsec:msi}

User intent is revealed not only at different temporal granularities but also through heterogeneous behavioral channels. We therefore organize the request-time input into three parts: a \textit{primary intent sequence} of live-room entries, \textit{auxiliary behavior sequences} such as likes, cart clicks, and orders, and \textit{long-term user profiles} containing profile and persistent-intent features. This separation allows the model to capture broad exploratory intent with dense behavioral coverage, deeper purchase-oriented intent with sparser but higher-confidence actions, and the user's stable preference prior.

Live-room entry forms the primary intent sequence because it is much denser than downstream purchase-oriented behaviors and provides broad coverage for request-time intent modeling after ad exposure. Let $\mathbf{a}^{\text{entry}}(t)=[\mathbf{a}^{\text{entry}}_1,\ldots,\mathbf{a}^{\text{entry}}_L]$ denote the most recent entry actions available at request time $t$. MSI applies strides $\mathcal{S}=\{1,2,10\}$ to construct fine-, intermediate-, and coarse-grained entry tokens. Fine-grained tokens preserve rapid short-term intent shifts, whereas coarser tokens summarize longer-range behavioral patterns; together with the long-term profile, they capture both the user's immediate intent and stable preferences. For each stride $s\in\mathcal{S}$:

\begin{equation*}
\mathbf{T}_{\text{entry},j}^{(s)}(t) = \mathbf{W}_s \left[ \mathbf{a}^{\text{entry}}_{s(j-1)+1}; \dots; \mathbf{a}^{\text{entry}}_{sj} \right] + \mathbf{p}_j^{(s)}, \quad j = 1, \dots, \left\lfloor \frac{L}{s} \right\rfloor.
\end{equation*}
The resulting sequences form the multi-scale representation of primary intent:

\begin{equation*}
\mathbf{T}_{\text{entry}}^{\text{multi}}(t) = \left[ \mathbf{T}_{\text{entry}}^{(s_1)}(t); \mathbf{T}_{\text{entry}}^{(s_2)}(t); \mathbf{T}_{\text{entry}}^{(s_3)}(t) \right].
\end{equation*}

Entry actions capture the user's immediate willingness to explore a live room, whereas auxiliary behaviors characterize intent depth and item-level preference: likes indicate content recognition, cart clicks indicate purchase consideration, and orders indicate completed purchase commitment. We denote these auxiliary streams as
$\mathcal{A}^{\text{aux}}=\{\mathbf{a}^{\text{like}},\mathbf{a}^{\text{cart}},\mathbf{a}^{\text{order}},\ldots\}$. These streams differ substantially in sparsity and temporal pattern. TAGR therefore preserves their action semantics by encoding each stream separately, rather than merging all actions into a single chronological history. The resulting behavior-specific token sequences are concatenated with the multi-scale entry representation and long-term profile tokens:

\begin{equation*}
\mathbf{M}_u(t) = \mathrm{LN} \left( \left[ \mathbf{T}_{\text{entry}}^{\text{multi}}(t); \mathbf{T}_{\text{like}}(t); \mathbf{T}_{\text{cart}}(t); \mathbf{T}_{\text{order}}(t);...; \mathbf{T}_{\text{profile}} \right] \right).
\end{equation*}

The resulting $\mathbf{M}_u(t)$ exposes both the behavioral scale and behavior type directly to the decoder. Attention can therefore select not only the temporal horizon most relevant to the current request, but also the action channel that best expresses the user's current decision stage. MSI thus forms a time-dependent, multi-scale, multi-behavior representation of request-time intent, anchored by the dense entry sequence and refined by deeper auxiliary behavior streams. As new behaviors enter these sequences, refreshing $\mathbf{M}_u(t)$ adapts the representation without discarding the coarse-grained patterns and profile features that encode stable preference.

\subsubsection{Multi-Factor Weighted Next-Token Prediction}
\label{subsec:mf_ntp}

Multi-scale conditioning represents intent using information available at request time, but it does not determine how reliably the subsequent logged target expresses that intent. In live-stream advertising, a brief exposure or entry provides only weak evidence that the displayed live ad matches the user's underlying product intent, whereas downstream actions such as cart clicks and orders provide progressively stronger evidence. Treating all logged targets uniformly allows frequent but weak signals to dominate next-token learning. We therefore weight each logged target according to \textit{behavior-depth intent evidence}, \textit{user value tier}, and \textit{commercial value}. Specifically, we assign each training sample $x$ a multiplicative weight:

\begin{equation*}
w(x) = \underbrace{w_{\text{feedback}}(x)}_{\text{Intent Evidence}} \cdot \underbrace{w_{\text{user}}(x) \cdot w_{\text{eCPM}}(x)}_{\text{Business Value}}.
\end{equation*}

\begin{itemize}
    \item \textbf{Intent evidence $w_{\text{feedback}}(x)$:} The intent evidence weight uses post-request behavior depth as intent evidence that the logged target matches the user's intent. Samples followed by deeper feedback, such as cart clicks or orders, receive more likelihood mass than those with only shallow exposure or entry feedback.
    \item \textbf{Business value:} The user-value-tier term $w_{\text{user}}(x)$ assigns a stratum-specific base weight using the user's long-term advertising-value label $u_{\text{type}}$. The commercial-value term $w_{\text{eCPM}}(x)$ maps eCPM into a normalized quantile space and increases the contribution of samples in higher intervals; quantile normalization prevents extreme values from dominating optimization.
\end{itemize}

These factors play distinct roles. The behavior-depth intent evidence weight estimates how reliably the logged target reflects user intent, whereas the user value tier and commercial value terms determine how much value weight the sample should receive in advertising recommendation; none of them is intended to estimate a calibrated outcome probability. MSI thus determines \textit{which request-time behavioral signals} represent intent, and $w_{\text{feedback}}$ determines \textit{how confidently} the logged target supervises that intent. Since LSID generation proceeds from coarse to fine levels, we additionally introduce level-wise factors $\gamma=[\gamma_1,\dots,\gamma_D]$. The MF-NTP objective is:
\begin{equation}
\mathcal{L}_{\text{NTP}} = - \frac{\sum_{x \in \mathcal{B}} w(x) \sum_{\ell=1}^{D} \gamma_\ell \cdot \log P_\theta(s_\ell^x | s_{<\ell}^x, \mathbf{M}_u^x(t_x))}{\sum_{x \in \mathcal{B}} w(x) \cdot \sum_{\ell=1}^{D} \gamma_\ell}
\label{eq:ntp_loss}
\end{equation}

where $\mathcal{B}$ is a training batch and $t_x$ is the request time of sample $x$. In implementation, the sample weights are normalized and clipped for robustness. Post-request real-user feedback is used only to weight supervision during training; serving-time generation depends solely on the request-time context $\mathbf{M}_u(t)$ and does not require future feedback. Together, MSI and MF-NTP construct intent from multi-scale, multi-behavior histories and learn it from logged targets weighted by behavior-depth intent evidence and business value.

\begin{figure}[tbp]
        \centering
        \includegraphics[width=1.0\linewidth]{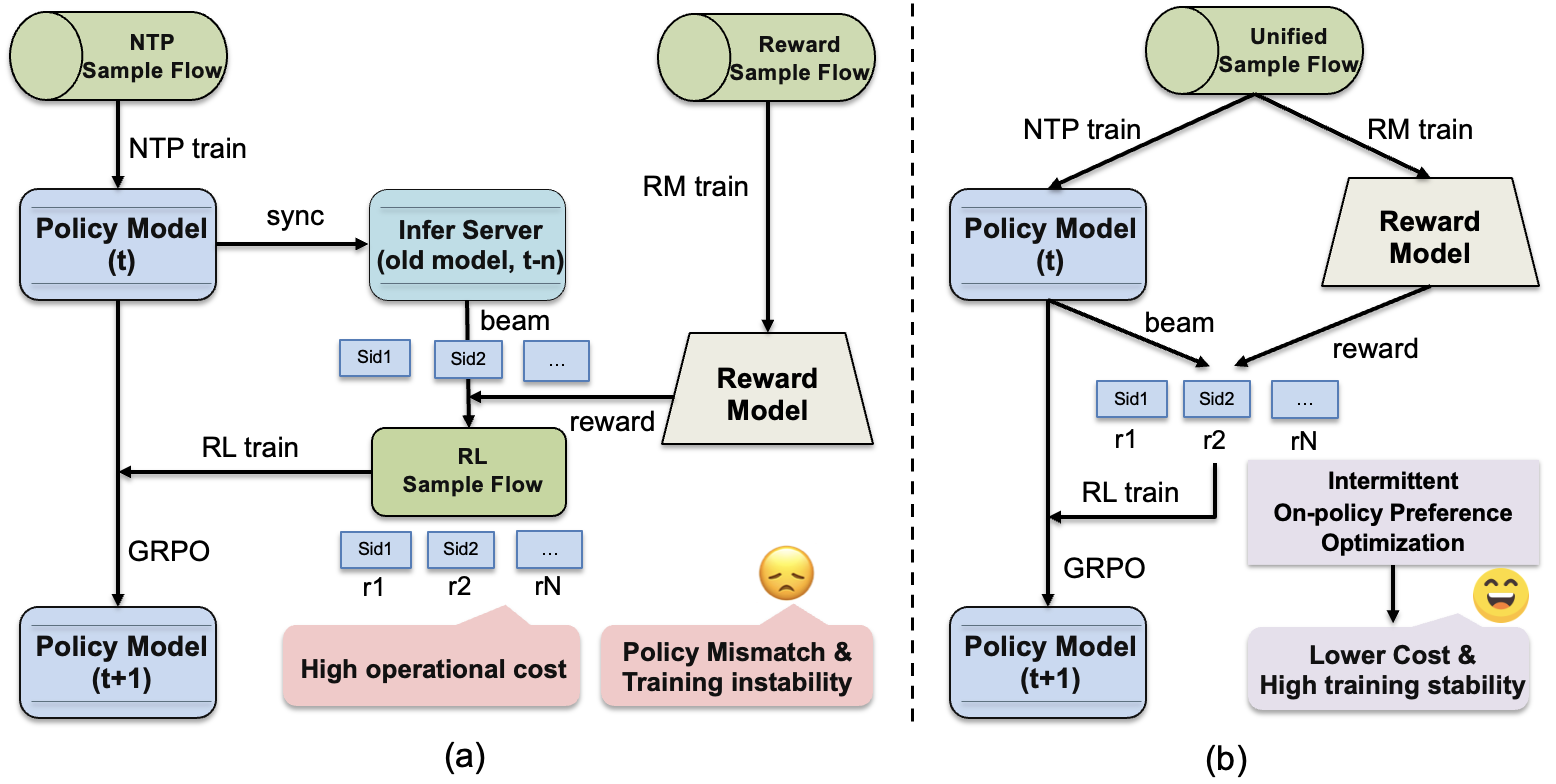}
\vspace{-2em}
    \caption{Preference alignment in SID-based generative recommendation systems: (a) Typical \emph{off-policy} preference optimization learns from a lagged-policy candidate pool, whereas (b) IOPO performs \emph{on-policy} candidate generation and applies intermittent preference updates while jointly training the Reward Model.}
    \label{fig:system}
\vspace{-1em}
\end{figure}


\subsection{Intermittent On-Policy Preference Optimization}
\label{subsec:iopo}


At the preference alignment level, reliable online adaptation depends on feedback that remains representative of the current policy's candidate distribution. Traditional SID-based generative recommendation systems~\cite{onerec,gr4ad} typically perform \emph{off-policy} preference optimization, relying on candidates generated by a lagged policy; similar off-policy effects have also been observed in efficient RL training pipelines~\cite{yao2025offpolicy}. Consequently, their supervision may become stale as the current policy and traffic distribution move away from the collection policy.
As shown in Figure~\ref{fig:system}, this off-policy paradigm differs from IOPO in that training candidates are not produced by the current generator being optimized. While such a design supports scalable deployment, it introduces two limitations in live-stream advertising:
\begin{itemize}
    \item \textbf{Policy mismatch and training instability.} Candidates generated by $\pi_{\text{old}}$ gradually deviate from the current policy $\pi_\theta$ as the generator and traffic distribution evolve. The resulting distribution mismatch makes reward supervision biased and noisy, which can destabilize training in rapidly changing live-stream scenarios.
    \item \textbf{High operational cost.} Operating the sample server incurs substantial overhead for SID generation, reward computation, and pool management. When multiple policy versions iterate concurrently, each version requires a separate sample pool, further compounding resource demand.
\end{itemize}



To address these challenges, we propose \textbf{Intermittent On-Policy Preference Optimization (IOPO)}, a plug-and-play RL preference optimization method integrated into our streaming online training strategy. In TAGR, newly collected real-user feedback is used to update both the NTP model and the RL policy online. IOPO organizes this online learning into two sequential phases. During \textbf{warmup}, we jointly train the NTP module and Reward Model (RM) to obtain a reliable base policy $\pi_\theta$ and a stable value estimator. During \textbf{intermittent on-policy adaptation}, the current policy periodically generates fresh candidate groups, followed by short bursts of group-relative policy optimization (GRPO) updates every $T$ steps; intervening online NTP maintenance steps continue learning from logged user feedback, anchor the generator to the observed behavior distribution, and mitigate catastrophic forgetting~\cite{ewc}. Each GRPO burst combines Behavior-Aligned GRPO (BA-GRPO), which preserves faithfulness to user behavior, with Value-Aligned GRPO (VA-GRPO), which incorporates downstream commercial utility. By replacing continuous online RL with intermittent preference updates, IOPO retains current-policy feedback while reducing reward-gradient interference with supervised NTP and improving training stability.

\subsubsection{Behavior-Aligned GRPO (BA-GRPO)}

BA-GRPO provides the behavioral anchor needed during current-policy adaptation. It encourages generated LSIDs to remain close to the user's next observed interaction in the learned token embedding space, preventing value optimization from drifting away from actual user behavior. We use a lightweight on-policy GRPO variant that directly weights each current-policy sequence log-likelihood by its clipped, group-normalized advantage.

Let $\hat{s}^{(k)} = [\hat{s}_1^{(k)}, \dots, \hat{s}_D^{(k)}]$ denote the $k$-th generated LSID sequence, and $s^* = [s_1^*, \dots, s_D^*]$ the ground-truth LSID, which is the same next-LSID label used in supervised NTP (i.e., the next observed interaction mapped to an LSID under the same labeling pipeline). We embed both sequences by summing level-wise token embeddings and define:

\begin{align}
\hat{\mathbf{e}}_k &= \sum_{\ell=1}^{D} \mathbb{E}_\ell(\hat{s}_\ell^{(k)}), \quad
\mathbf{e}^* = \sum_{\ell=1}^{D} \mathbb{E}_\ell(s_\ell^*) \label{eq:ba_embed} \\
r_k^{\text{ba}} &= \frac{1 + \cos(\hat{\mathbf{e}}_k, \mathbf{e}^*)}{2} \in [0,1] \label{eq:ba_reward} \\
\hat{A}_k^{\text{ba}} &= \operatorname{clip}\left( \frac{r_k^{\text{ba}} - \mu^{\text{ba}}}{\sigma^{\text{ba}}+\epsilon}, \; c_{\text{low}}, \; c_{\text{high}} \right) \label{eq:ba_adv} \\
\mathcal{L}_{\text{GRPO}}^{\text{ba}} &= -\frac{1}{|\mathcal{B}| G} \sum_{x \in \mathcal{B}} \sum_{k=1}^{G} w(x) \, \log \pi_\theta(\hat{s}^{(k)} \mid \mathbf{M}_u^x(t_x)) \, \hat{A}_k^{\text{ba}} \label{eq:ba_loss}
\end{align}
where $\mu^{\text{ba}}$ and $\sigma^{\text{ba}}$ are the mean and standard deviation within the $G$ candidates generated for the same request, $\epsilon$ is a numerical stabilizer, and $w(x)$ is the intent-and-utility sample weight in Eq.~\eqref{eq:ntp_loss}. The group-normalized advantage makes rewards comparable within each current-policy rollout.

\subsubsection{Value-Aligned GRPO (VA-GRPO)}

Behavioral similarity alone does not ensure that the generated set contains high-value candidates. VA-GRPO therefore complements the behavioral anchor with an RM that scores generated LSIDs using business-aware signals. The RM shares bottom-layer LSID embeddings with TAGR for representation consistency, while task-specific heads predict post-exposure live-room entry and normalized eCPM; in our system, the eCPM label is taken from the production fine-ranking score, and the post-exposure LRE label is the real user feedback after exposure. A stop-gradient is applied during RM scoring to keep reward estimation stable.

Let $\mathbf{c}_k$ denote the eCPM context features. The value-aligned reward is:

\begin{align}
r_k^{\text{post}} &= \sigma\bigl( \mathrm{MLP}_{\text{post}}([\bar{\mathbf{M}}_u(t_x); \hat{\mathbf{e}}_k]) \bigr) \label{eq:va_post} \\
r_k^{\text{eCPM}} &= \operatorname{clamp}\bigl( \mathrm{MLP}_{\text{eCPM}}([\bar{\mathbf{M}}_u(t_x); \hat{\mathbf{e}}_k; \mathbf{c}_k]), 0, 1 \bigr) \label{eq:va_ecpm} \\
r_k^{\text{va}} &= r_k^{\text{post}} + \beta_{\text{va}} \cdot r_k^{\text{eCPM}} \label{eq:va_reward} \\
\hat{A}_k^{\text{va}} &= \operatorname{clip}\left( \frac{r_k^{\text{va}} - \mu^{\text{va}}}{\sigma^{\text{va}}+\epsilon}, \; c_{\text{low}}, \; c_{\text{high}} \right) \label{eq:va_adv} \\
\mathcal{L}_{\text{GRPO}}^{\text{va}} &= -\frac{1}{|\mathcal{B}| G} \sum_{x \in \mathcal{B}} \sum_{k=1}^{G} w(x)\,\log \pi_\theta(\hat{s}^{(k)} \mid \mathbf{M}_u^x(t_x)) \, \hat{A}_k^{\text{va}} \label{eq:va_loss}
\end{align}

BA-GRPO and VA-GRPO optimize the same current policy but play complementary roles. BA-GRPO supplies a parameter-free behavior signal that preserves relevance to the user's next interaction, whereas VA-GRPO steers generation toward candidates with stronger business value.
Here, $\mu^{\text{va}}$ and $\sigma^{\text{va}}$ are computed within the current-policy candidate group, analogously to BA-GRPO.

\subsubsection{Combined GRPO Loss}

The final preference objective combines the two GRPO losses:
\begin{equation}
\mathcal{L}_{\text{GRPO}} = \lambda_{\text{ba}}\, \mathcal{L}_{\text{GRPO}}^{\text{ba}} + \lambda_{\text{va}}\, \mathcal{L}_{\text{GRPO}}^{\text{va}}
\label{eq:combined_loss}
\end{equation}
where $\lambda_{\text{ba}}$ and $\lambda_{\text{va}}$ balance behavior and value alignment.

The overall training objective applies the GRPO loss only during scheduled RL steps:
\begin{equation}
\mathcal{L} = \mathcal{L}_{\text{NTP}} + \mathbb{I}_{\text{RL}} \cdot \mathcal{L}_{\text{GRPO}},
\label{eq:joint_loss}
\end{equation}
where $\mathbb{I}_{\text{RL}}$ equals $1$ during scheduled on-policy steps and $0$ otherwise. IOPO thus realizes alignment-level temporal adaptation: scheduled current-policy updates keep preference feedback fresh, while intervening supervised maintenance prevents reward optimization from continuously overwriting the learned behavior distribution.

\section{System Deployment}
\label{sec:deployment}

\subsection{Online Training Platform}

\begin{figure}[tbp]
        \centering
        \includegraphics[width=1.0\linewidth]{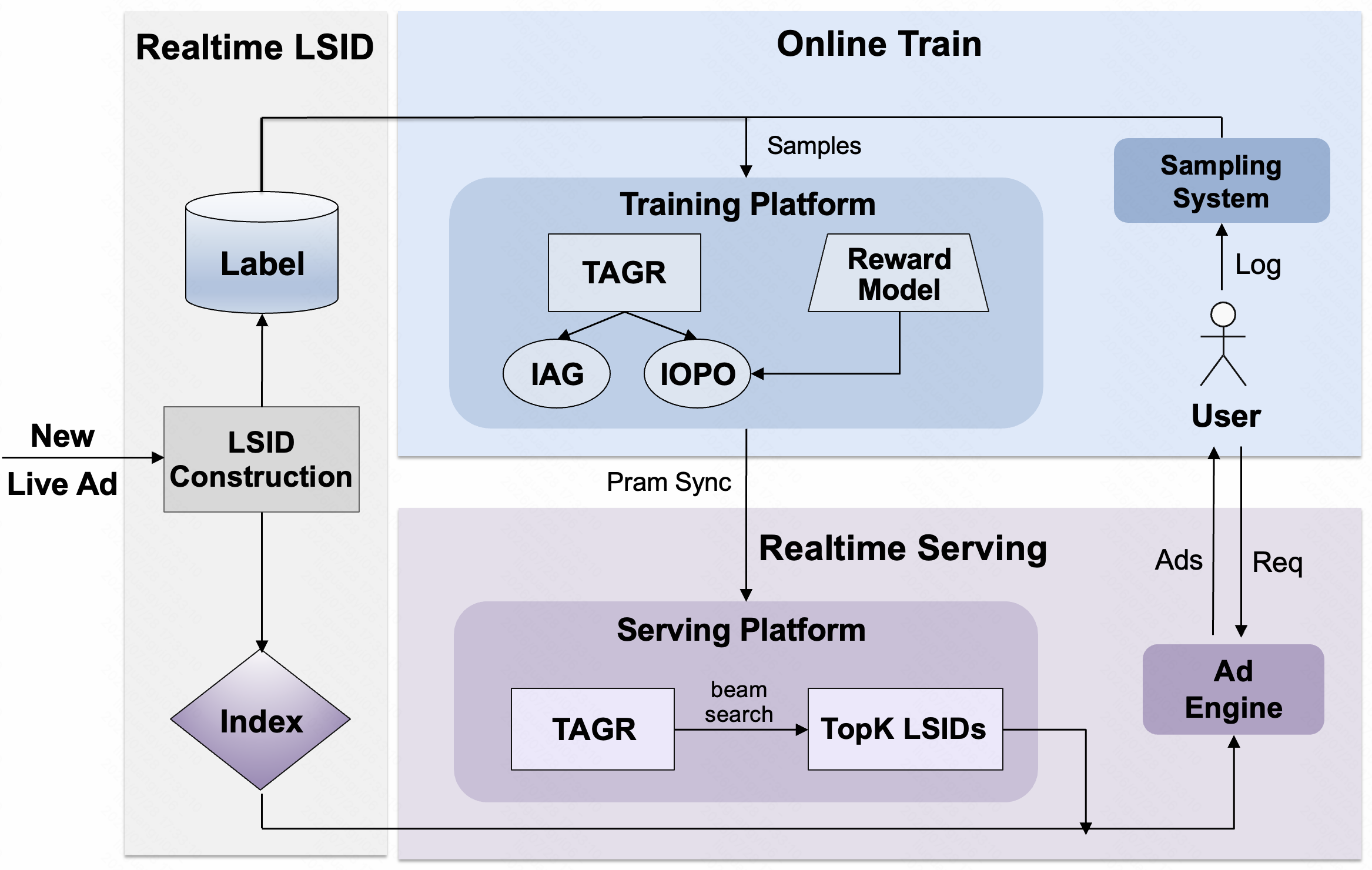}
\vspace{-2em}
    \caption{System deployment: training and serving of TAGR.}
    \label{fig:deploy}
\vspace{-1em}
\end{figure}

We have deployed TAGR as a generative retrieval in Kuaishou's live-stream advertising system, serving more than 400 million daily active users. As shown in Figure~\ref{fig:deploy}, the closed-loop system connects real-time LSID construction, online training, and real-time serving.

The platform processes streaming request logs and interaction feedback to update IAG and IOPO. IAG jointly trains MSI-encoder and live decoder to model user intent and balance intent evidence with business value (Sections~\ref{subsec:msi} and~\ref{subsec:mf_ntp}). IOPO periodically samples current-policy candidates and uses the reward model (RM) to estimate engagement and commercial value; intermittent updates preserve stable preference alignment in the non-stationary live-stream environment (Section~\ref{subsec:iopo}).

TAGR, including its generator and RM, is updated from the same streaming data pipeline with separate task-specific objectives. The generator and RM share bottom-layer LSID embeddings for representation consistency, while a stop-gradient is applied during RM scoring to keep reward estimation stable. The platform performs continual mini-batch training and periodically synchronizes updated parameters to the serving cluster.

\subsection{Real-time LSID Construction}

When a new live ad becomes active or its scene and promoted products change, the LSID Construction module computes its LSID as described in Section~\ref{subsec:lsid}; active live ads are re-encoded every minute. The refreshed assignment is immediately written to online LSID storage as the canonical token label and propagated to the bidirectional LSID--LiveID index within seconds. Because only token assignments are refreshed while the hierarchical vocabulary remains fixed, new and evolving live ads become retrievable without resource-intensive global index rebuilding.

\subsection{Real-time Serving Engine}

Upon receiving a request, the Real-time Serving Engine performs beam search with the latest TAGR decoder to generate the Top-$K_{\mathrm{serve}}$ LSIDs ($K_{\mathrm{serve}}=256$), resolves them to active live ads through the LSID--LiveID index, and passes the candidates to downstream filtering and ranking. Optimized decoding and result caching sustain more than 2500 QPS per L20 GPU with end-to-end retrieval latency below 100\,ms. The request, generated LSIDs, user feedback, and fine-ranking signals such as eCPM are logged back to the Online Training Platform, completing the closed loop.

\section{Experiments}

\subsection{Experimental Setup}


\begin{table*}[t]
\centering
\caption{Overall performance comparison. Offline recall is evaluated by HR@K on the live-streaming dataset for generative retrieval methods; online metrics are measured by A/B tests on production traffic under the same downstream filtering/ranking stack and comparable traffic buckets. Best results are bolded. All lifts are relative to DLRM (base). LRE = live-room entry, SCC = shopping-cart click. LRE-Rate Lift and SCC-Rate Lift denote relative improvements in post-exposure rate.}
\label{tab:overall_performance}
\begin{tabular}{l *{7}{c}}
\toprule
\multirow{2}{*}{Method}
  & \multicolumn{2}{c}{\textbf{LRE} (Offline)}
  & \multicolumn{2}{c}{\textbf{SCC} (Offline)}
  & \multicolumn{3}{c}{\textbf{Online (A/B)}} \\
\cmidrule(lr){2-3} \cmidrule(lr){4-5} \cmidrule(lr){6-8}
  & HR@64 & HR@128 & HR@64 & HR@128 & LRE-Rate Lift & SCC-Rate Lift & Revenue Lift \\
\midrule
DLRM (base) & \textemdash & \textemdash & \textemdash & \textemdash & Ref. & Ref. & Ref. \\
OneRec v2 (GR base)  & 0.5945 & 0.6568  & 0.5649 & 0.6216  & +2.1\% & +1.8\% & +6.1\% \\
\midrule
TAGR w/ LSID  & 0.6413 & 0.7024  & 0.6058 & 0.6551  & +4.2\% & +3.6\% & +9.9\% \\
\quad + IAG (MSI-Encoding)  & 0.6678 & 0.7276  & 0.6174 & 0.6709  & +5.1\% & +4.5\% & +11.2\% \\
\quad + IAG (MF-NTP)  & 0.6871 & 0.7468  & 0.6319 & 0.6836  & +6.7\% & +6.0\% & +13.5\% \\
\quad + IOPO (BA-GRPO) & 0.7026 & 0.7604  & 0.6397 & 0.6901  & +7.5\% & +6.6\% & +14.7\% \\
\quad + IOPO (VA-GRPO)  & \textbf{0.7135} & \textbf{0.7723}  & \textbf{0.6461} & \textbf{0.6965}  & \textbf{+8.5\%} & \textbf{+7.4\%} & \textbf{+16.1\%} \\
\bottomrule
\end{tabular}
\end{table*}


\subsubsection{Dataset and Evaluation Metrics}
We conduct experiments on a billion-scale dataset collected from a real-world e-commerce live-stream advertising platform. We use a temporal split comprising 5 days of training logs, the subsequent 2 days for validation, and the following 2 days for testing, thereby evaluating forward-looking generalization under rapid content drift. To avoid temporal leakage, all post-request feedback used for sample weighting and target construction is strictly restricted to the corresponding data split. The dataset contains behavior logs from over 400 million users and hundreds of thousands of live ads (at the live ad granularity), covering the full conversion funnel from exposure to live-room entry, shopping-cart clicks, and purchases. For each request, the user behavior sequences are built from the most recent 30 days of interactions.

\textbf{Offline metrics.} We evaluate retrieval quality using Hit Rate at multiple cutoffs (HR@K) for two critical engagement actions: live-room entry (\textbf{LRE}, \texttt{live\_room\_entry}) and shopping-cart click (\textbf{SCC}, \texttt{shopping\_cart\_click}). For each action, we construct supervised examples by taking the live ad associated with the user's \emph{next} occurrence of that action as the single ground-truth target (requests without a subsequent target action are excluded under a fixed post-request window, consistently applied to all methods). Following the standard protocol for generative retrieval, HR@K is computed by resolving the model-generated Top-$K_{\text{serve}}$ LSIDs to active live ads and matching them against the ground-truth target under the full candidate space. HR@K is a useful offline proxy because it directly measures whether the truly engaged live ad appears in the limited Top-$K_{\text{serve}}$ retrieval set, a prerequisite for downstream ranking and monetization; in our production experiments, relative HR@K improvements are directionally consistent with online behavior-rates and revenue lifts.

\textbf{Online metrics.} We conduct an \textit{online A/B test} on production traffic and report two complementary business metrics: (1) \textbf{Behavior Rate Lift (\%)}, which measures the relative improvement in post-exposure live-room entry rate (\textbf{LRE-Rate}) and shopping-cart click rate (\textbf{SCC-Rate}) over the DLRM baseline; and (2) \textbf{Revenue Lift (\%)}, which measures the relative improvement in realized advertising revenue and captures end-to-end commercial effectiveness.

\subsubsection{Baseline Methods}

We compare against \textbf{DLRM (base)}, a two-tower multi-channel discriminative retrieval baseline implemented on the DLRM architecture~\cite{dlrm}, and \textbf{OneRec v2 (GR base)}, a strong industrial generative recommendation baseline~\cite{onerec_v2}. 
DLRM uses fixed-ID ANN retrieval across multiple channels; since its offline HR@K reflects ranking quality over the full candidate corpus---different from autoregressive Top-$K$ LSID decoding where candidates are directly generated---the two metrics are not directly comparable and HR@K for DLRM is therefore not reported. 
OneRec v2 is deployed as a production generative retrieval baseline for online comparison. 
All methods share the same data budget and infrastructure constraints.

\subsection{Overall Performance}

Table~\ref{tab:overall_performance} presents a comprehensive comparison of all methods across offline recall and online business metrics.

\textbf{Generative vs.\ discriminative retrieval.} DLRM is the online two-tower multi-channel retrieval baseline and provides the reference point for production A/B lifts, while its offline HR@K is not directly comparable with generative Top-$K$ LSID decoding. OneRec v2 serves as the deployed generative retrieval baseline and already improves online business metrics over DLRM. TAGR further improves both offline recall and online business metrics by replacing static target identifiers with LSIDs, modeling temporal user intent, and performing intermittent current-policy preference optimization. These results demonstrate the benefit of jointly modeling live-stream dynamics and advertising value within an end-to-end generative retriever.

\textbf{Progressive gains from three-level temporal adaptation.} Each stage contributes consistent and additive improvements. At the token level, replacing static identifiers with LSID raises revenue lift to 9.9\% by refreshing the generation target with the current live scene and promoted products. At the intent level, MSI-Encoding increases the lift to 11.2\% through multi-scale, multi-behavior user conditioning, and MF-NTP further reaches 13.5\% by combining post-request intent signal with user and commercial value weighting. At the alignment level, BA-GRPO and VA-GRPO raise the lift to 14.7\% and 16.1\%, respectively, showing the complementary effects of behavior and value alignment under IOPO's intermittent current-policy updates. This progression directly supports TAGR's token-, intent-, and alignment-level decomposition.

\textbf{Full TAGR.} The complete model with all components achieves state-of-the-art results across all metrics, yielding \textbf{+8.5\%} and \textbf{+7.4\%} relative lifts in LRE-Rate and SCC-Rate, respectively, and a \textbf{+16.1\%} Revenue Lift over the DLRM baseline. Notably, the revenue lift substantially exceeds the behavior-rate improvements, indicating that TAGR not only retrieves more engaging ads but also surfaces higher-value inventory. This result is consistent with the commercial-value weighting in MF-NTP and the value-aligned reward in VA-GRPO.

\subsection{Online A/B Test}
\label{subsec:online_deployment}


We conduct a multi-week randomized experiment on 10\% of production traffic, covering more than 40 million users, with realized advertising revenue as the primary metric. Relative to the production DLRM, TAGR improves LRE rate by 8.5\%, SCC rate by 7.4\%, and revenue by \textbf{16.1\%}. Segment analysis shows a \textbf{28.8\%} revenue lift among low-value users and an \textbf{18.4\%} lift for cold-start live streams. These results are consistent with the broader mid-to-tail coverage in Figure~\ref{fig:recall}: token-side refresh improves access to newly active inventory, multi-scale and multi-behavior encoding tracks request-time intent, and MF-NTP together with IOPO prioritize candidates supported by stronger intent signals and advertising value.


\subsection{Ablation Studies}


\subsubsection{Token-Side Adaptation: LSID}

\begin{table}[tbp]
\centering
\small
\caption{Ablation of LSID along two dimensions: target construction and codebook size. ``+H'' denotes streamer-ID hashing at the final hierarchical level (see Section~\ref{subsec:lsid}).}
\label{tab:lsid_ablation}
\resizebox{0.99\columnwidth}{!}{%
\begin{tabular}{lcccc}
\toprule
\textbf{Token Settings} & \multicolumn{4}{c}{\textbf{Metrics}} \\
\midrule
\textit{Token Construction} & \textbf{Cpr}$\downarrow$ & \textbf{Col}$\downarrow$ & \textbf{Stability}$\uparrow$ & \textbf{Revenue}$\uparrow$ \\[2pt]
\quad Static SID (baseline) & 1.42 & 0.42 & 100\% & - \\
\quad Dynamic SID & 1.36 & 0.39 & 92\% & +1.7\% \\
\quad Dynamic SID + H & 1.09 & 0.15 & 92\% & +2.5\% \\
\quad LSID (full, +H) & \textbf{1.01} & \textbf{0.02} & 90\% & \textbf{+3.8\%} \\
\midrule
\textit{Codebook Size} & \textbf{Cpr}$\downarrow$ & \textbf{Col}$\downarrow$ & \textbf{Stability}$\uparrow$ & \textbf{Revenue}$^{\dagger}\uparrow$ \\[2pt]
\quad [256,256,256] & 1.02 & 0.02 & 90\% & - \\
\quad [128,128,128] & 1.14 & 0.16 & {93\%} & -1.4\% \\
\quad [512,512,512] & {1.01} & {0.01} & 86\% & {+0.12\%} \\
\bottomrule
\end{tabular}
}
\vspace{-10pt}
\end{table}

Table~\ref{tab:lsid_ablation} reports four metrics. \textbf{Cpr} (compression ratio$\downarrow$) is the number of live-ad items divided by the number of distinct LSIDs assigned to them, and therefore measures the average number of items represented by each LSID; the ideal value is 1.0. \textbf{Col} (collision rate$\downarrow$) is the fraction of distinct LSIDs that are assigned to multiple items rather than mapping one-to-one. \textbf{Stability}$\uparrow$ measures the proportion of live ads whose level-1 assignment remains unchanged throughout a live session, reflecting target consistency during serving; and \textbf{Revenue} reports online revenue lift. Revenue lifts in the \textit{Target Construction} block are relative to Static SID, whereas those in the \textit{Codebook Size} block are relative to LSID with codebook size [256,256,256]. All refreshed variants maintain at least 90\% level-1 stability, indicating that token-side temporal adaptation remains sufficiently stable for production serving.

\textbf{Token-construction evolution.} Each upgrade reduces the compression ratio towards the ideal value of 1.0. A lower Cpr means that each generated identifier maps to fewer live ads in the serving index, making candidate expansion more targeted and efficient. Dynamic SID periodically updates each live ad's semantic representation and token assignment as its content evolves but does not incorporate collaborative signals from user–scene–product alignment. This semantic-only dynamic update lowers Cpr from 1.42 to 1.36 and yields a 1.7\% revenue gain. Adding streamer-ID hashing (+H) further lowers Cpr to 1.09 and raises the gain to 2.5\%. Finally, full LSID incorporates user–scene–product collaborative alignment into the dynamically updated scene–product representation, achieving near-ideal compression (Cpr=1.01), the lowest collision rate, and the largest revenue lift of 3.8\%. These results show that dynamic semantic updates and collaborative alignment provide complementary benefits.

\textbf{Codebook size trade-off.} With the LSID construction fixed, varying the codebook size primarily affects compression and serving costs. A smaller codebook [128,\allowbreak 128,\allowbreak 128] over-compresses the target space (Cpr=1.14), forcing dissimilar live ads to share identifiers; each generated LSID consequently expands to a larger and noisier candidate list, degrading revenue by 1.4\%. A larger codebook [512,512,512] achieves near-ideal compression (Cpr=1.01), but its larger vocabulary increases decoding costs for only a marginal revenue gain of 0.12\%. The default [256,256,256] achieves the same near-ideal compression at a lower serving cost, offering the best revenue-efficiency trade-off.

\begin{table}[t]
\centering
\small
\caption{Fine-grained ablation of Multi-Scale Intent-Aware Generation.}
\label{tab:intent_generation_ablation}
\resizebox{0.99\columnwidth}{!}{%
\begin{tabular}{lcc}
\toprule
\textbf{Intent-Side Settings} & \multicolumn{2}{c}{\textbf{Metrics}} \\
\midrule
\textit{MSI-Encoding: Scales} & \textbf{LRE HR@128} & \textbf{SCC HR@128} \\[2pt]
\quad Entry stride $=\{1\}$ (baseline)       & 0.7024  & 0.6551  \\
\quad Entry strides $=\{1,2\}$               & 0.7141  & 0.6638 \\
\quad Entry strides $=\{1,2,10\}$            & 0.7276  & 0.6709 \\
\midrule
\textit{MF-NTP: Weighting Factors} & \textbf{LRE HR@128} & \textbf{SCC HR@128} \\[2pt]
\quad Uniform NTP (baseline)                                  & 0.7276 & 0.6709 \\
\quad $+w_{\text{feedback}}$                           & 0.7352 & 0.6757 \\
\quad $+w_{\text{feedback}}+w_{\text{user}}$           & 0.7405 & 0.6789 \\
\quad Full MF-NTP ($+w_{\text{eCPM}}$)                        & \textbf{0.7468} & \textbf{0.6836} \\
\bottomrule
\end{tabular}
}
\vspace{-8pt}
\end{table}

\subsubsection{Intent-Side Adaptation: IAG}

\label{subsec:intent_generation_ablation}

Table~\ref{tab:intent_generation_ablation} provides a fine-grained analysis of how MSI-Encoding represents request-time intent and how MF-NTP weights the corresponding logged targets.

\paragraph{MSI-Encoding Ablation.}
We vary the temporal granularities of the primary live-room-entry sequence (as defined in Section~\ref{subsec:msi}). Stride 1 captures the most recent entries and short-term intent shifts, stride 2 captures local exploration patterns, and stride 10 summarizes longer-range browsing intent. As shown in the first block of Table~\ref{tab:intent_generation_ablation}, progressively incorporating these temporal scales improves both live-room-entry and shopping-cart-click retrieval. The consistent gains indicate that fine-grained recent actions and coarser browsing patterns provide complementary evidence for request-time intent.

\paragraph{MF-NTP Ablation.}
The second block of Table~\ref{tab:intent_generation_ablation} introduces the weighting factors in the same conceptual order as the method. Starting from uniform NTP, the intent-evidence weight $w_{\text{feedback}}$ uses post-request behavior depth as evidence that the logged live ad matches the user's intent, while the user-value and eCPM terms further adjust the training priority according to long-term user value and normalized commercial value. Their progressive inclusion yields sustained performance improvements, supporting the separation of \textit{intent signal} from \textit{value weighting} rather than treating heterogeneous outcomes as a single undifferentiated importance signal.

\subsubsection{Alignment-Side Adaptation: IOPO}

We dissect IOPO from three angles: (1) on-policy vs.\ off-policy candidates and update frequency; (2) the behavior-alignment contribution of BA-GRPO; and (3) the reward dynamics of VA-GRPO. We further compare preference-optimization backbones in Appendix~\ref{app:preference_backbone}.

\paragraph{On-Policy vs.\ Off-Policy and Update Frequency.}
As shown in Table~\ref{tab:freq}, we compare three update strategies across two axes: candidate source and update frequency. \textbf{Off-policy GRPO} follows prior off-policy alignment schemes~\cite{gr4ad}, drawing candidates from a fixed lagged policy $\pi_{\text{old}}$; the resulting distribution gap yields the lowest reward and poorest stability. Among the current-policy variants, \textbf{Continuous GRPO} removes this gap but suffers from repeated reward-gradient interference with supervised learning. \textbf{Intermittent GRPO (IOPO)} achieves the best NTP loss (1.53), reward (0.913), and stability at only one-third of the training cost of continuous updates.

\begin{table}[tbp]
\centering
\setlength{\tabcolsep}{3pt}
\caption{Ablation of on-policy update strategy. Stability = proportion of training windows without loss spikes ($>$0.1 above running mean).}
\label{tab:freq}
\resizebox{\columnwidth}{!}{%
\begin{tabular}{lcccc}
\toprule
\textbf{Method} & \textbf{NTP Loss $\downarrow$} & \textbf{Reward $\uparrow$} & \textbf{Stability $\uparrow$} & \textbf{LRE HR@128} \\
\midrule
\textit{Off-policy} \\
\quad GRPO & 1.79  & 0.675  & low & 0.6982 \\
\midrule
\textit{On-policy} \\
\quad Continuous GRPO   & 1.68          & 0.864         & moderate & 0.7469 \\
\quad Intermittent GRPO (IOPO) & \textbf{1.53} & \textbf{0.913} & \textbf{high} & \textbf{0.7723}  \\
\bottomrule
\end{tabular}
}
\end{table}

\begin{figure}[tbp]
        \centering
        \includegraphics[width=1.0\linewidth]{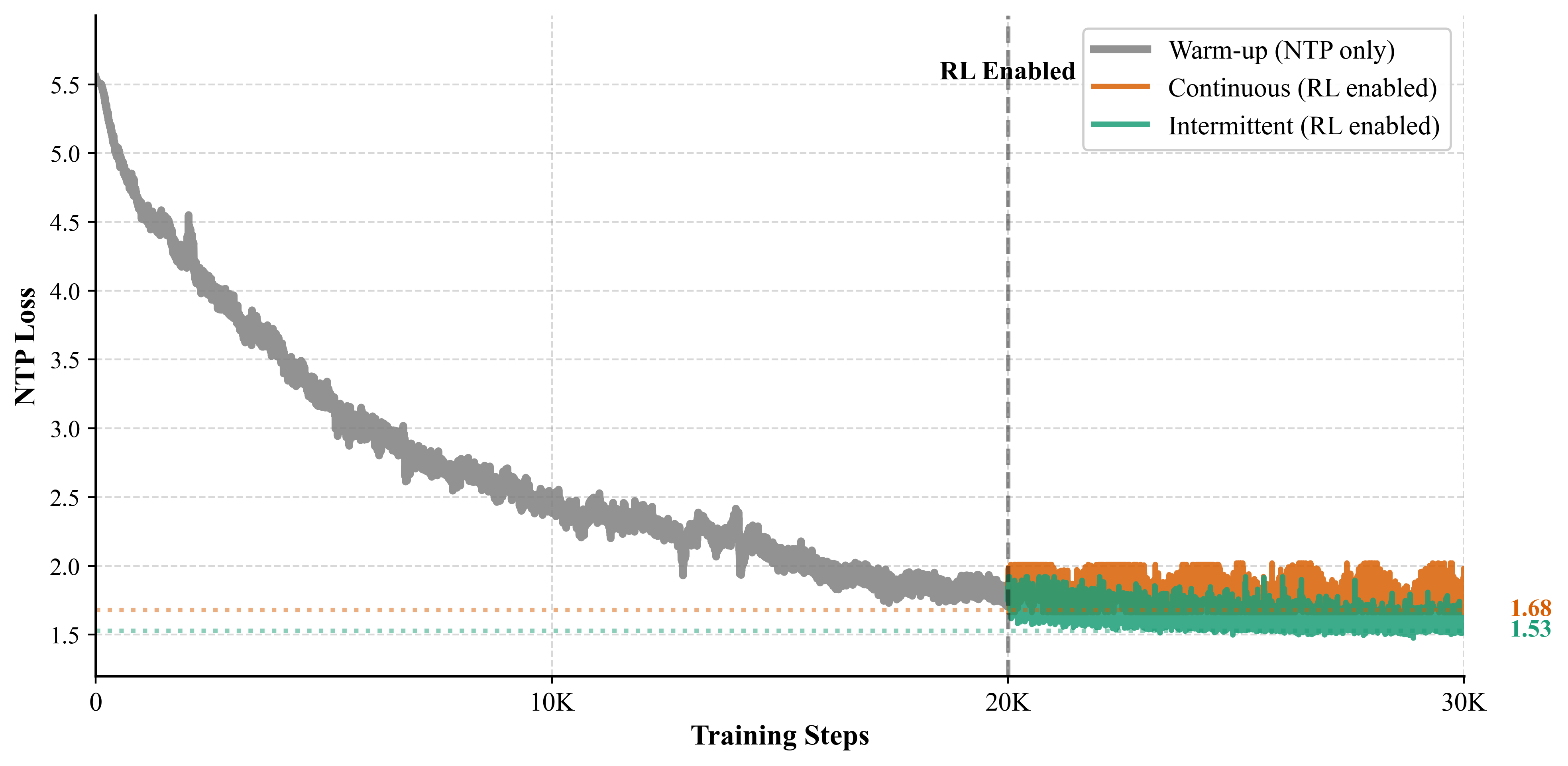}
\vspace{-2em}
    \caption{NTP Loss: Continuous vs.\ Intermittent RL Updates.}
    \label{fig:intermittent}
\vspace{-1em}
\end{figure}

Figure~\ref{fig:intermittent} zooms into the on-policy frequency comparison. During warm-up (0--20K steps), loss drops from $\sim$5.5 to $\sim$1.7. At step 20K, RL is activated: continuous learning (orange) immediately begins oscillating with a large amplitude (ranging $\sim$1.5--2.0) and stagnates at $\sim$1.68, reflecting persistent gradient interference with the NTP signal. Intermittent learning (green), by alternating RL bursts with NTP maintenance steps, produces significantly smaller oscillations and settles at $\sim$1.53. The off-policy variant is omitted from the figure as its unstable trajectory obscures the on-policy comparison; its final metrics are reported in Table~\ref{tab:freq}.

\paragraph{Behavior-Aligned GRPO.}
Table~\ref{tab:ba_grpo_ablation} isolates the two ingredients of BA-GRPO. Starting from the supervised Multi-Scale Intent-Aware Generation model, unweighted BA-GRPO introduces current-policy optimization using token-space similarity to the user's next observed interaction. Reusing the MF-NTP sample weight $w(x)$ further accounts for the reliability of the post-request intent signal and the value priority of each sample. The progressive gains show that behavioral alignment benefits from both a request-level similarity reward and sample-level supervision strength.

\begin{table}[tbp]
\centering
\caption{Ablation of behavior alignment and MF-NTP sample weighting in BA-GRPO.}
\label{tab:ba_grpo_ablation}
\resizebox{\columnwidth}{!}{%
\begin{tabular}{lcc}
\toprule
\textbf{Method} & \textbf{LRE HR@128} & \textbf{SCC HR@128} \\
\midrule
Supervised IAG & 0.7468 & 0.6836 \\
+BA-GRPO (w/o sample weighting) & 0.7501 & 0.6858 \\
+BA-GRPO (with MF-NTP weighting) & \textbf{0.7604} & \textbf{0.6901} \\
\bottomrule
\end{tabular}
}
\end{table}

\begin{figure}[tbp]
        \centering
        \includegraphics[width=1.0\linewidth]{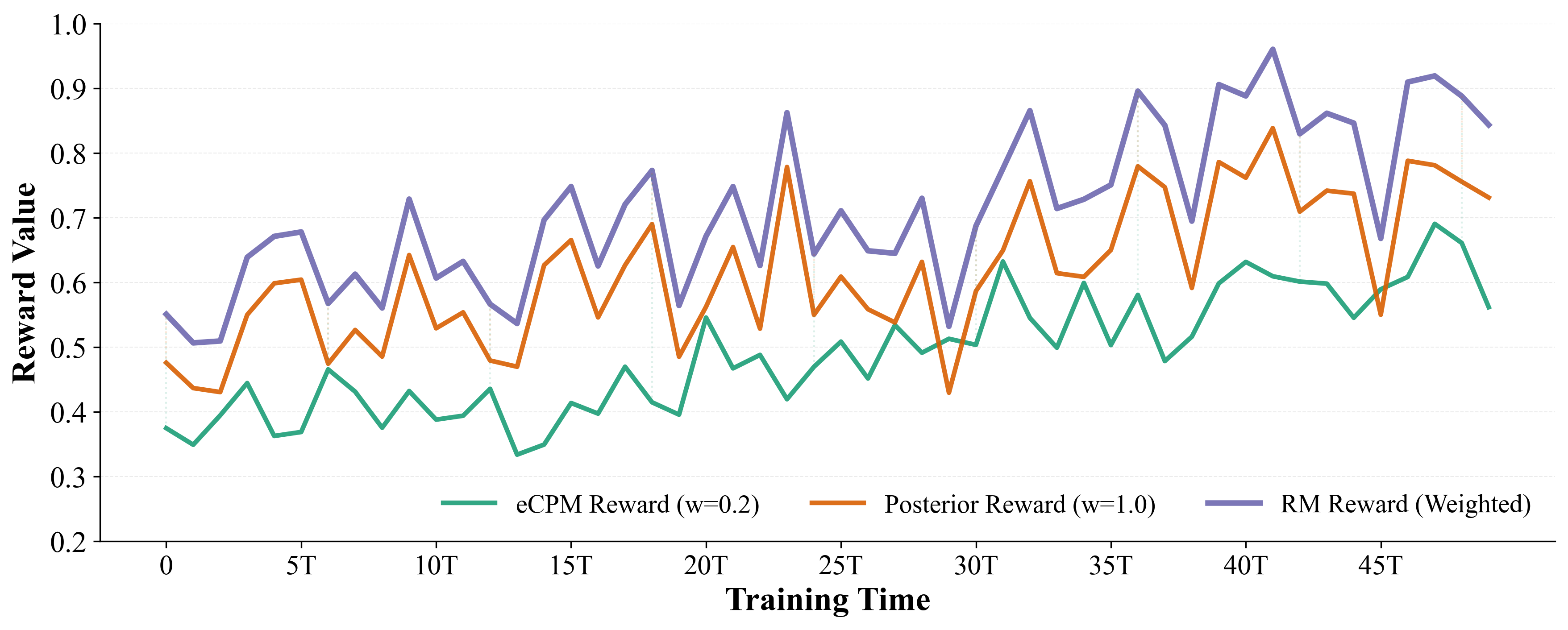}
\vspace{-2em}
    \caption{Training dynamics of the reward components in VA-GRPO.}
    \label{fig:reward}
\vspace{-1em}
\end{figure}

\paragraph{Value-Aligned GRPO.}
Figure~\ref{fig:reward} illustrates the training dynamics of the post-exposure response score $r^{\text{post}}$, normalized commercial-value score $r^{\text{eCPM}}$, and their value-aligned combination $r^{\text{va}}$. All three curves exhibit a consistent upward trend, showing that current-policy optimization progressively shifts candidate generation toward stronger engagement and commercial outcomes. Following Eq.~\eqref{eq:va_reward}, VA-GRPO uses $r^{\text{va}}=r^{\text{post}}+\beta_{\text{va}}r^{\text{eCPM}}$: the response component supplies the primary engagement signal, while the weighted eCPM component incorporates commercial value without overwhelming behavioral relevance.

\section{Related Work}

\subsection{Live-Streaming Recommendation}
Live-streaming recommendation must react to rapidly changing content, short item lifecycles, and delayed or heterogeneous user feedback~\cite{liverec,contentctr,mmbee}. Discriminative studies improve freshness and real-time modeling through sliding-window streams, cross-domain live signals, and future behavior and content prediction~\cite{sliver,momentcross,tsstfn,farm,larm,liveforesighter,foresight}. More recently, generative models have been introduced into this scenario. OneLive~\cite{onelive} employs dynamic tokenization and time-aware modeling for end-to-end live-stream generation, and SSRLive~\cite{ssrlive} combines dynamic SIDs with user--streamer interaction signals in a generative--discriminative framework. Existing work primarily targets general live recommendation; TAGR instead treats each live ad as a dynamic scene--product target and couples refreshed LSIDs with intent- and value-aware learning.

\subsection{Generative Recommendation}
Generative recommendation reformulates retrieval as autoregressive generation over discrete item tokens, replacing large-vocabulary ID classification or ANN retrieval with structured decoding. TIGER \cite{tiger} establishes a representative semantic-ID paradigm by quantizing item representations into hierarchical codes. Subsequent work enriches the generated token space and user conditioning with collaborative and semantic information~\cite{letter,eager,mmq_v2}. LC-Rec~\cite{lcrec} integrates collaborative semantics into language-model-based recommendation, OneRec~\cite{onerec} unifies retrieval and ranking in an end-to-end generative architecture, and Align$^3$GR~\cite{align3gr} performs token-, behavior-, and preference-level alignment for LLM-based generative recommendation. In industrial advertising, GR4AD~\cite{gr4ad} develops a production-oriented generative recommender with ad-aware tokenization, efficient decoding, and value-oriented learning. Most non-live generative recommenders nevertheless assume stable target semantics; TAGR instead refreshes each live ad's request-independent LSID as its scene or product changes while preserving a stable vocabulary.

\subsection{Preference Optimization}
Preference optimization complements supervised next-token learning by directly shaping generated candidates according to comparative user or value signals. DPO~\cite{dpo} provides a simple offline objective for learning from pairwise preferences without an explicit policy-gradient loop, and S-DPO~\cite{sdpo} adapts this idea to recommendation by incorporating multiple negatives and ranking-aware softmax structure. Recent generative recommenders further integrate preference alignment into end-to-end systems: OneRec~\cite{onerec} constructs iterative preference data for DPO, Align$^3$GR~\cite{align3gr} combines self-play and real-feedback DPO for progressive alignment, and GR4AD~\cite{gr4ad} introduces a list-wise value-oriented preference objective for advertising. Group Relative Policy Optimization (GRPO)~\cite{grpo} offers another alternative by normalizing rewards within a group of sampled candidates and avoiding a separate critic. Existing alignment faces stale off-policy feedback or unstable continuous updates; IOPO balances freshness and stability through intermittent current-policy GRPO interleaved with NTP maintenance.

\section{Conclusion}

We present \textbf{TAGR}, a temporally adaptive generative recommendation framework for industrial live-stream advertising. TAGR addresses temporal variation at three complementary levels. LSID refreshes token assignments as live scenes and promoted products change while preserving a stable hierarchical token vocabulary. IAG constructs a time-dependent user representation from multi-scale behavior encoding; its MF-NTP objective further weights supervision by the reliability of post-request intent evidence and by business value. IOPO incorporates fresh current-policy feedback through intermittent preference updates while using supervised NTP maintenance to stabilize online learning. Experiments on a large-scale live-stream advertising platform show that TAGR outperforms discriminative and generative baselines, including a \textbf{+16.1\%} revenue lift in production. These results establish temporal adaptation as an effective design principle for generative retrieval in non-stationary industrial environments.

\bibliographystyle{ACM-Reference-Format}
\bibliography{sample-base}

@inproceedings{tiger,
  title     = {Recommender Systems with Generative Retrieval},
  author    = {Rajput, Shashank and Mehta, Nikhil and Singh, Anima and Keshavan, Raghunandan Hulikal and Vu, Trung and Heldt, Lukasz and Hong, Lichan and Tay, Yi and Tran, Vinh Q. and Samost, Jonah and Kula, Maciej and Chi, Ed H. and Sathiamoorthy, Maheswaran},
  booktitle = {Advances in Neural Information Processing Systems},
  year      = {2023}
}

@article{onerec,
  title   = {OneRec: Unifying Retrieve and Rank with Generative Recommender and Iterative Preference Alignment},
  author  = {Deng, Jiaxin and Wang, Shiyao and Cai, Kuo and Ren, Lejian and Hu, Qigen and Ding, Weifeng and Luo, Qiang and Zhou, Guorui},
  journal = {arXiv preprint arXiv:2502.18965},
  year    = {2025}
}

@article{onerec_v2,
  title   = {OneRec-V2 Technical Report},
  author  = {Zhou, Guorui and Hu, Hengrui and Cheng, Hongtao and Wang, Huanjie and Deng, Jiaxin and Zhang, Jinghao and Cai, Kuo and Ren, Lejian and Ren, Lu and Yu, Liao and others},
  journal = {arXiv preprint arXiv:2508.20900},
  year    = {2025}
}

@article{gr4ad,
  title   = {Generative Recommendation for Large-Scale Advertising},
  author  = {Xue, Ben and Liu, Dan and Wang, Lixiang and Sun, Mingjie and Wang, Peng and Zhang, Pengfei and Shi, Shaoyun and Xu, Tianyu and Sha, Yunhao and Liu, Zhiqiang and Kong, Bo and Wang, Bo and Yang, Hang and Xue, Jieting and Wang, Junhao and Wang, Shengyu and Hui, Shuping and Ye, Wencai and Lin, Xiao and Li, Yongzhi and Chen, Yuhang and Yin, Zhihui and Chen, Quan and Wen, Shiyang and Wu, Wenjin and Li, Han and Zhou, Guorui and Li, Changcheng and Jiang, Peng},
  journal = {arXiv preprint arXiv:2602.22732},
  year    = {2026}
}

@inproceedings{das,
  title={DAS: Dual-Aligned Semantic IDs Empowered Industrial Recommender System},
  author={Ye, Wencai and Sun, Mingjie and Shi, Shaoyun and Wang, Peng and Wu, Wenjin and Jiang, Peng},
  booktitle={Proceedings of the 34th ACM International Conference on Information and Knowledge Management},
  pages={6217--6224},
  year={2025}
}

@inproceedings{lcrec,
  title     = {Adapting Large Language Models by Integrating Collaborative Semantics for Recommendation},
  author    = {Zheng, Bowen and Hou, Yupeng and Lu, Hongyu and Chen, Yu and Zhao, Wayne Xin and Chen, Ming and Wen, Ji-Rong},
  booktitle = {2024 IEEE 40th International Conference on Data Engineering (ICDE)},
  pages     = {1435--1448},
  year      = {2024},
  doi       = {10.1109/ICDE60146.2024.00118}
}

@inproceedings{align3gr,
  title     = {Align$^3$GR: Unified Multi-Level Alignment for LLM-based Generative Recommendation},
  author    = {Ye, Wencai and Sun, Mingjie and Chen, Shuhang and Wu, Wenjin and Jiang, Peng},
  booktitle = {Proceedings of the AAAI Conference on Artificial Intelligence},
  volume    = {40},
  number    = {19},
  pages     = {16154--16162},
  year      = {2026},
  doi       = {10.1609/aaai.v40i19.38651}
}

@article{openonerec,
  title   = {OpenOneRec Technical Report},
  author  = {Zhou, Guorui and Bao, Honghui and Huang, Jiaming and Deng, Jiaxin and Zhang, Jinghao and She, Junda and Cai, Kuo and Ren, Lejian and Ren, Lu and Luo, Qiang and Wang, Qianqian and Hu, Qigen and Zhang, Rongzhou and Tang, Ruiming and Wang, Shiyao and others},
  journal = {arXiv preprint arXiv:2512.24762},
  year    = {2025}
}

@article{grpo,
  title   = {{DeepSeekMath}: Pushing the Limits of Mathematical Reasoning in Open Language Models},
  author  = {Shao, Zhihong and Wang, Peiyi and Zhu, Qihao and Xu, Runxin and Song, Junxiao and Bi, Xiao and Zhang, Haowei and Zhang, Mingchuan and Li, Y. K. and Wu, Y. and Guo, Daya},
  journal = {arXiv preprint arXiv:2402.03300},
  year    = {2024}
}

@inproceedings{transformer,
  title     = {Attention Is All You Need},
  author    = {Vaswani, Ashish and Shazeer, Noam and Parmar, Niki and
               Uszkoreit, Jakob and Jones, Llion and Gomez, Aidan N. and
               Kaiser, {\L}ukasz and Polosukhin, Illia},
  booktitle = {Advances in Neural Information Processing Systems},
  volume    = {30},
  pages     = {5998--6008},
  year      = {2017}
}

@article{infonce,
  title   = {Representation Learning with Contrastive Predictive Coding},
  author  = {van den Oord, A{\"a}ron and Li, Yazhe and Vinyals, Oriol},
  journal = {arXiv preprint arXiv:1807.03748},
  year    = {2018}
}

@inproceedings{rqvae,
  title     = {Autoregressive Image Generation Using Residual Quantization},
  author    = {Lee, Doyup and Kim, Chiheon and Kim, Saehoon and
               Cho, Minsu and Han, Wook-Shin},
  booktitle = {Proceedings of the IEEE/CVF Conference on Computer Vision and Pattern Recognition},
  pages     = {11513--11522},
  year      = {2022},
  doi       = {10.1109/CVPR52688.2022.01123}
}

@article{ewc,
  title   = {Overcoming Catastrophic Forgetting in Neural Networks},
  author  = {Kirkpatrick, James and Pascanu, Razvan and Rabinowitz, Neil and
             Veness, Joel and Desjardins, Guillaume and Rusu, Andrei A. and
             Milan, Kieran and Quan, John and Ramalho, Tiago and
             Grabska-Barwinska, Agnieszka and Hassabis, Demis and
             Clopath, Claudia and Kumaran, Dharshan and Hadsell, Raia},
  journal = {Proceedings of the National Academy of Sciences},
  volume  = {114},
  number  = {13},
  pages   = {3521--3526},
  year    = {2017},
  doi     = {10.1073/pnas.1611835114}
}

@inproceedings{din,
  title     = {Deep Interest Network for Click-Through Rate Prediction},
  author    = {Zhou, Guorui and Zhu, Xiaoqiang and Song, Chengru and
               Fan, Ying and Zhu, Han and Ma, Xiao and Yan, Yanghui and
               Jin, Junqi and Li, Han and Gai, Kun},
  booktitle = {Proceedings of the 24th ACM SIGKDD International Conference on Knowledge Discovery \& Data Mining},
  pages     = {1059--1068},
  year      = {2018},
  doi       = {10.1145/3219819.3219823}
}

@inproceedings{sim,
  title     = {Search-Based User Interest Modeling with Lifelong Sequential Behavior Data for Click-Through Rate Prediction},
  author    = {Pi, Qi and Zhou, Guorui and Zhang, Yujing and Wang, Zhe and
               Ren, Lejian and Fan, Ying and Zhu, Xiaoqiang and Gai, Kun},
  booktitle = {Proceedings of the 29th ACM International Conference on Information \& Knowledge Management},
  pages     = {2685--2692},
  year      = {2020},
  doi       = {10.1145/3340531.3412744}
}

@article{dlrm,
  title   = {Deep Learning Recommendation Model for Personalization and Recommendation Systems},
  author  = {Naumov, Maxim and Mudigere, Dheevatsa and Shi, Hao-Jun Michael and
             Huang, Jianyu and Sundaraman, Narayanan and Park, Jongsoo and
             Wang, Xiaodong and Gupta, Udit and Wu, Carole-Jean and
             Azzolini, Alisson G. and Dzhulgakov, Dmytro and Mallevich, Andrey and
             Cherniavskii, Ilia and Lu, Yinghai and Krishnamoorthi, Raghuraman and
             Yu, Ansha and Kondratenko, Volodymyr and Pereira, Stephanie and
             Chen, Xianjie and Chen, Wenlin and Rao, Vijay and Jia, Bill and
             Xiong, Liang and Smelyanskiy, Misha},
  journal = {arXiv preprint arXiv:1906.00091},
  year    = {2019}
}

@article{sliver,
  title   = {Ensure Timeliness and Accuracy: A Novel Sliding Window Data Stream Paradigm for Live Streaming Recommendation},
  author  = {Liang, Fengqi and Zheng, Baigong and Zhao, Liqin and Zhou, Guorui and Wang, Qian and Niu, Yanan},
  journal = {arXiv preprint arXiv:2402.14399},
  year    = {2024}
}

@article{momentcross,
  title   = {Moment\&Cross: Next-Generation Real-Time Cross-Domain CTR Prediction for Live-Streaming Recommendation at Kuaishou},
  author  = {Cao, Jiangxia and Wang, Shen and Li, Yue and Wang, Shenghui and Tang, Jian and Wang, Shiyao and Yang, Shuang and Liu, Zhaojie and Zhou, Guorui},
  journal = {arXiv preprint arXiv:2408.05709},
  year    = {2024}
}

@article{liveforesighter,
  title   = {LiveForesighter: Generating Future Information for Live-Streaming Recommendations at Kuaishou},
  author  = {Lu, Yucheng and Cao, Jiangxia and Kuan, Xu and Cheng, Wei and Jiang, Wei and Zhang, Jiaming and Yang, Shuang and Liu, Zhaojie and Hong, Liyin},
  journal = {arXiv preprint arXiv:2502.06557},
  year    = {2025}
}

@article{onelive,
  title   = {OneLive: Dynamically Unified Generative Framework for Live-Streaming Recommendation},
  author  = {Wang, Shen and Huang, Yusheng and Yang, Ruochen and Wen, Shuang and Xu, Pengbo and Cao, Jiangxia and Liu, Yueyang and Cai, Kuo and others},
  journal = {arXiv preprint arXiv:2602.08612},
  year    = {2026}
}

@article{ssrlive,
  title   = {SSRLive: Live Streaming Recommendation with Dynamic Semantic ID},
  author  = {Shi, Teng and Li, Zhaoheng and Qu, Yuanhang and Liu, Yi and Lai, Lixiang and Jiang, Yuning},
  journal = {arXiv preprint arXiv:2606.06970},
  year    = {2026}
}

@inproceedings{dpo,
  title     = {Direct Preference Optimization: Your Language Model Is Secretly a Reward Model},
  author    = {Rafailov, Rafael and Sharma, Archit and Mitchell, Eric and Manning, Christopher D. and Ermon, Stefano and Finn, Chelsea},
  booktitle = {Advances in Neural Information Processing Systems},
  volume    = {36},
  year      = {2023}
}

@inproceedings{sdpo,
  title     = {On Softmax Direct Preference Optimization for Recommendation},
  author    = {Chen, Yuxin and Tan, Junfei and Zhang, An and Yang, Zhengyi and Sheng, Leheng and Zhang, Enzhi and Wang, Xiang and Chua, Tat-Seng},
  booktitle = {Advances in Neural Information Processing Systems},
  volume    = {37},
  year      = {2024}
}

@inproceedings{liverec,
  title     = {Recommendation on Live-Streaming Platforms: Dynamic Availability and Repeat Consumption},
  author    = {Rappaz, J{\'e}r{\'e}mie and McAuley, Julian and Aberer, Karl},
  booktitle = {Proceedings of the 15th ACM Conference on Recommender Systems},
  pages     = {390--399},
  year      = {2021},
  doi       = {10.1145/3460231.3474267}
}

@article{contentctr,
  title   = {{ContentCTR}: Frame-Level Live Streaming Click-Through Rate Prediction with Multimodal Transformer},
  author  = {Deng, Jiaxin and Shen, Dong and Wang, Shiyao and Wu, Xiangyu and Yang, Fan and Zhou, Guorui and Meng, Gaofeng},
  journal = {arXiv preprint arXiv:2306.14392},
  year    = {2023}
}

@inproceedings{mmbee,
  title     = {{MMBee}: Live Streaming Gift-Sending Recommendations via Multi-Modal Fusion and Behaviour Expansion},
  author    = {Deng, Jiaxin and Wang, Shiyao and Wang, Yuchen and Qi, Jiansong and Zhao, Liqin and Zhou, Guorui and Meng, Gaofeng},
  booktitle = {Proceedings of the 30th ACM SIGKDD Conference on Knowledge Discovery and Data Mining},
  pages     = {4896--4905},
  year      = {2024},
  doi       = {10.1145/3637528.3671511}
}

@inproceedings{tsstfn,
  title     = {A Bilateral Perspective for Modeling Real-Time Traffic Trends in Live-Streaming Recommendation},
  author    = {Li, Rui and Gao, Pengyuan and Li, Haihan and Chai, Ling and Huang, Shaohao and Xie, Ting},
  booktitle = {2025 IEEE 41st International Conference on Data Engineering (ICDE)},
  pages     = {4142--4155},
  year      = {2025},
  publisher = {IEEE}
}

@article{farm,
  title   = {{FARM}: Frequency-Aware Model for Cross-Domain Live-Streaming Recommendation},
  author  = {Li, Xiaodong and Yang, Ruochen and Wen, Shuang and Wang, Shen and Liu, Yueyang and Wang, Guoquan and Hu, Weisong and Luo, Qiang and Sheng, Jiawei and Liu, Tingwen and others},
  journal = {arXiv preprint arXiv:2502.09375},
  year    = {2025}
}

@article{larm,
  title   = {{LLM}-Alignment Live-Streaming Recommendation},
  author  = {Liu, Yueyang and Cao, Jiangxia and Wang, Shen and Wen, Shuang and Chen, Xiang and Wu, Xiangyu and Yang, Shuang and Liu, Zhaojie and Gai, Kun and Zhou, Guorui},
  journal = {arXiv preprint arXiv:2504.05217},
  year    = {2025}
}

@article{foresight,
  title   = {Foresight Prediction Enhanced Live-Streaming Recommendation},
  author  = {Cao, Jiangxia and Yang, Ruochen and Chen, Xiang and Lao, Changxin and Liu, Yueyang and Huang, Yusheng and Tian, Yuanhao and Wu, Xiangyu and Yang, Shuang and Liu, Zhaojie and others},
  journal = {arXiv preprint arXiv:2512.06700},
  year    = {2025}
}

@inproceedings{letter,
  title     = {Learnable Item Tokenization for Generative Recommendation},
  author    = {Wang, Wenjie and Bao, Honghui and Lin, Xinyu and Zhang, Jizhi and Li, Yongqi and Feng, Fuli and Ng, See-Kiong and Chua, Tat-Seng},
  booktitle = {Proceedings of the 33rd ACM International Conference on Information and Knowledge Management},
  pages     = {2400--2409},
  year      = {2024},
  doi       = {10.1145/3627673.3679569}
}

@inproceedings{eager,
  title     = {{EAGER}: Two-Stream Generative Recommender with Behavior-Semantic Collaboration},
  author    = {Wang, Ye and Xun, Jiahao and Hong, Minjie and Zhu, Jieming and Jin, Tao and Lin, Wang and Li, Haoyuan and Li, Linjun and Xia, Yan and Zhao, Zhou and Dong, Zhenhua},
  booktitle = {Proceedings of the 30th ACM SIGKDD Conference on Knowledge Discovery and Data Mining},
  pages     = {3245--3254},
  year      = {2024},
  doi       = {10.1145/3637528.3671775}
}

@article{mmq_v2,
  title   = {{MMQ-v2}: Align, Denoise, and Amplify: Adaptive Behavior Mining for Semantic {ID} Learning in Recommendation},
  author  = {Xu, Yi and Zhang, Moyu and Fan, Chaofan and Hu, Jinxin and Li, Xiaochen and Zhang, Yu and Zeng, Xiaoyi and Zhang, Jing},
  journal = {arXiv preprint arXiv:2510.25622},
  year    = {2025}
}

@misc{yao2025offpolicy,
  title = {Your Efficient RL Framework Secretly Brings You Off-Policy RL Training},
  url = {https://fengyao.notion.site/off-policy-rl},
  author = {Yao, Feng and Liu, Liyuan and Zhang, Dinghuai and Dong, Chengyu and Shang, Jingbo and Gao, Jianfeng},
  journal = {Feng Yao's Notion},
  year = {2025},
  month = aug,
}

\clearpage
\appendix
\section{Implementation Details}
\label{app:implementation}

For reproducibility, this section reports implementation details omitted from the main text, including the model architecture, data and sequence settings, training and serving hyperparameters, and the hyperparameter-selection protocol.


\subsection{Model Architecture and Hyperparameters}

Table~\ref{tab:hyperparams} summarizes the Lazy Decoder architecture and the main hyperparameters used in training and serving. The decoder follows a lightweight Transformer design with cross-attention over the encoded user context and autoregressive self-attention over the generated LSID prefix. General training settings are shared by the supervised NTP stage and IOPO stage unless otherwise specified.

\begin{table}[htbp]
\centering
\small
\caption{Model architecture and key training and serving hyperparameters.}
\label{tab:hyperparams}
\begin{tabular}{lcc}
\toprule
\textbf{Hyperparameter} & \textbf{Symbol} & \textbf{Value} \\
\midrule
\multicolumn{3}{l}{\textit{Lazy Decoder Architecture}} \\
Transformer decoder layers & -- & 6 \\
Hidden dimension & -- & 512 \\
Attention heads & -- & 4 \\
Feed-forward dimension & -- & 2048 \\
Activation & -- & ReLU \\
Dropout rate & -- & 0.1 \\
\midrule
\multicolumn{3}{l}{\textit{General Training}} \\
Optimizer & -- & Adam \\
Learning rate & $\eta$ & $1\times10^{-5}$ \\
Batch size & -- & 512 per GPU \\
Primary entry sequence length & $L$ & 300 \\
\midrule
\multicolumn{3}{l}{\textit{LSID -- Alignment Model}} \\
InfoNCE temperature & $\tau$ & 0.07 \\
Streamer ID hash bucket size & -- & 256 \\
Scene--Product fusion weights & $[\alpha, \beta]$ & $[0.8, 0.2]$ \\
LSID refresh cadence & -- & 1--5 min \\
\midrule
\multicolumn{3}{l}{\textit{MF-NTP}} \\
Hierarchical decay factors & $[\gamma_1, \gamma_2, \gamma_3]$ & $[1.0, 0.8, 0.6]$ \\
Sample weight clip threshold & $w_{\max}$ & 5.0 \\
\midrule
\multicolumn{3}{l}{\textit{IOPO -- BA-GRPO}} \\
Advantage clip range & $[c_{\text{low}}, c_{\text{high}}]$ & $[-2.5, 2.5]$ \\
Number of rollout candidates (group size) & $G$ & 20 \\
BA-GRPO loss weight & $\lambda_{\text{ba}}$ & 0.1 \\
\midrule
\multicolumn{3}{l}{\textit{IOPO -- VA-GRPO}} \\
Advantage clip range & $[c_{\text{low}}, c_{\text{high}}]$ & $[-2.5, 2.5]$ \\
eCPM reward weight & $\beta_{\text{va}}$ & 0.2 \\
VA-GRPO loss weight & $\lambda_{\text{va}}$ & 1.0 \\
\midrule
\multicolumn{3}{l}{\textit{IOPO Schedule}} \\
Intermittent update interval & $T$ & 200 \\
Warmup steps & -- & 20K \\
\midrule
\multicolumn{3}{l}{\textit{Serving}} \\
Beam search width & $K_{\text{serve}}$ & 256 \\
\bottomrule
\end{tabular}
\end{table}

\subsection{User Behavior Data and Sequence Settings}

\textbf{Primary entry sequence.} We truncate the primary live-room entry history to the most recent $L=300$ entry actions. MSI applies three strides $\mathcal{S}=\{1, 2, 10\}$ over this sequence, yielding subsequences of lengths $L$, $\lfloor L/2 \rfloor$, and $\lfloor L/10 \rfloor$, respectively. This multi-scale construction allows the model to capture recent entry intent, local exploration patterns, and longer-range browsing tendencies within the same request-time representation.

\textbf{Auxiliary behavior streams.} Auxiliary behavior streams, including product clicks, likes, shopping-cart clicks, and orders, are maintained and encoded separately from the primary entry sequence. This design preserves action semantics and avoids mixing dense exploratory signals with sparse but stronger purchase-oriented signals.


\subsection{Hyperparameter Selection}
\label{app:hparam_selection}

We tune the non-architectural hyperparameters on the held-out validation set comprising two days of offline data. The validation objective jointly considers LRE HR@128, SCC HR@128, and training stability. The search ranges include $\gamma_\ell \in \{1.0, 0.8, 0.6, 0.5\}$ for the level-wise NTP factors, $w_{\max} \in \{5, 10, 20\}$ for sample-weight clipping, $[c_{\text{low}}, c_{\text{high}}] \in \{[-2,2], [-2.5,2.5], [-3,3]\}$ for GRPO advantage clipping, $\lambda_{\text{ba}} \in \{0.1, 0.5, 1.0\}$, $\lambda_{\text{va}} \in \{0.5, 1.0\}$, $\beta_{\text{va}} \in \{0.1, 0.2, 0.5\}$, $[\alpha, \beta] \in \{[0.7,0.3], [0.8,0.2], [0.9,0.1]\}$, and $T \in \{100, 200, 500\}$. The selected values are reported in Table~\ref{tab:hyperparams}.

\section{Complementary Experiments}
\label{app:complementary_experiments}

This section provides two complementary analyses omitted from the main text: a comparison of preference-optimization backbones and an analysis of retrieval coverage across advertiser spend tiers.

\subsection{Preference Optimization Backbone}
\label{app:preference_backbone}

All three backbones are evaluated with the same intermittent current-policy rollouts, isolating the optimization algorithm from the candidate-freshness comparison in Table~\ref{tab:freq}. For a controlled comparison, they also share the same value model for scoring candidates and constructing preference signals. Let $G$ denote the candidate group size; in our default setting, $G$ equals the number of rollout candidates per request ($G=20$) unless otherwise stated. We compare DPO with one positive and one negative, Softmax-DPO with one positive and the remaining $G-1$ negatives, and GRPO with group-relative advantages over all $G$ candidates. Table~\ref{tab:dpo_grpo} shows that using more current-policy candidates consistently improves performance, while GRPO achieves the highest reward and retrieval accuracy. We therefore use GRPO as the optimization backbone for both BA-GRPO and VA-GRPO within IOPO.

\begin{table}[htbp]
\centering
\small
\caption{Algorithm comparison: preference optimization backbones.}
\label{tab:dpo_grpo}
\begin{tabular}{lccc}
\toprule
\textbf{Algorithm} & \textbf{Candidate Strategy} & \textbf{Reward $\uparrow$} & \textbf{LRE HR@128 $\uparrow$} \\
\midrule
DPO   & 1 pos / 1 neg (rank-$G$) & 0.834          & 0.7486          \\
S-DPO & 1 pos / $(G-1)$ neg       & 0.871          & 0.7594          \\
GRPO  & group of $G$              & \textbf{0.913} & \textbf{0.7723} \\
\bottomrule
\end{tabular}
\end{table}

\subsection{Retrieval Recall Distribution}

\begin{figure}[htbp]
        \centering
        \includegraphics[width=1.0\linewidth]{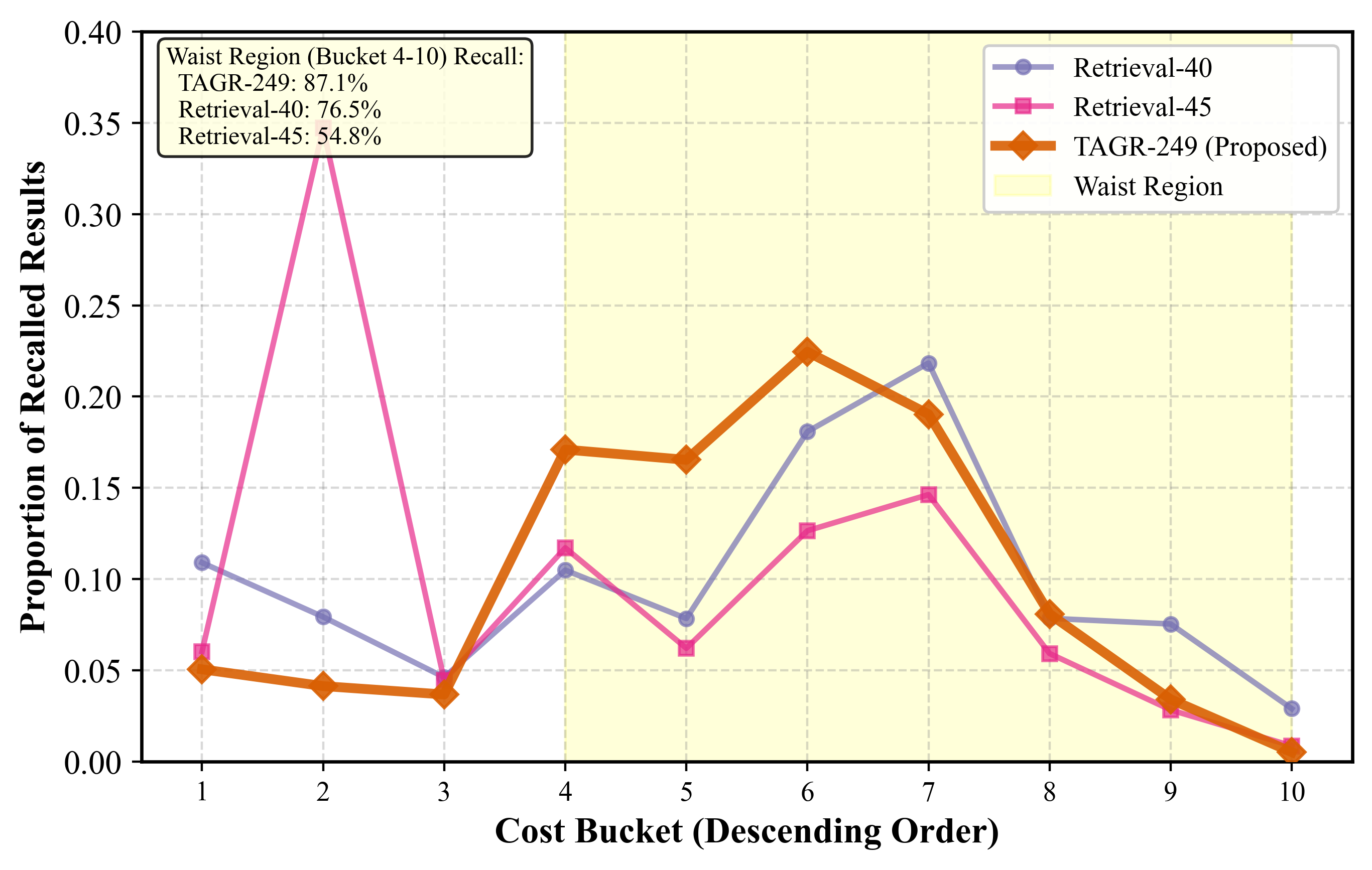}
\vspace{-2em}
    \caption{Comparison of recall distribution across cost tiers.}
    \label{fig:recall}
\vspace{-1em}
\end{figure}

To further investigate how different retrievers allocate their recall capacity across the advertiser spend spectrum, Fig.~\ref{fig:recall} reports the \emph{recall distribution} over streamer-level cost-consumption tiers. Specifically, we rank streamers by their cumulative advertising cost and partition them into buckets in descending order (Bucket~1: highest-spend streamers). For each request, we collect the Top-$K_{\text{serve}}$ recalled live ads from a given retrieval channel and map each recalled live ad to its streamer's bucket; we then aggregate over all requests and normalize the counts to obtain the proportion contributed by each bucket (y-axis). We compare two strong discriminative retrieval channels from the production system: \textbf{Retrieval-40} (ID 40, optimized towards live-room entry rate) and \textbf{Retrieval-45} (ID 45, optimized towards fine-ranking eCPM), against \textbf{TAGR-249} (our deployed TAGR channel, ID 249).

As illustrated, traditional discriminative models exhibit a pronounced \textit{top-tier concentration bias}, disproportionately allocating recall capacity to high-spend head buckets while underserving the broader waist region (Buckets 4--10). In contrast, TAGR-249 allocates substantially more recall to the waist region. This pattern is consistent with temporally refreshed LSID targets and intent-aware autoregressive generation, which enable the retriever to cover evolving inventory beyond the head ads favored by fixed-ID retrieval. The resulting shift toward broader mid-tier coverage is directionally consistent with TAGR's online gains reported in Section~\ref{subsec:online_deployment}.

\end{document}